\documentclass[twocolumn]{aastex62}

\usepackage{amsmath}

\shorttitle{Am Stars in LAMOST}
\shortauthors{Li Qin et al.}

\begin{document}
\bibliographystyle{aasjournal}

\title{Metallic-Line Stars Identified from Low Resolution Spectra of LAMOST DR5}

\correspondingauthor{A-Li Luo}
\email{lal@nao.cas.cn}

\author[0000-0001-6947-8684]{Li Qin}
\affiliation{Key Laboratory of Optical Astronomy, National Astronomical Observatories, Chinese Academy of Sciences,Beijing 100012, China}
\affiliation{School of Information Management \& Institute for Astronomical Science, Dezhou University, Dezhou 253023, China}
\affiliation{University of Chinese Academy of Sciences, Beijing 100049, China}

\author{A-Li Luo}
\affiliation{Key Laboratory of Optical Astronomy, National Astronomical Observatories, Chinese Academy of Sciences,Beijing 100012, China}
\affiliation{School of Information Management \& Institute for Astronomical Science, Dezhou University, Dezhou 253023, China}
\affiliation{University of Chinese Academy of Sciences, Beijing 100049, China}
\affiliation{Department of Physics and Astronomy, University of Delaware, Newark, DE 19716, USA}

\author{Wen Hou}
\affiliation{Key Laboratory of Optical Astronomy, National Astronomical Observatories, Chinese Academy of Sciences,Beijing 100012, China}

\author{Yin-Bi Li}
\affiliation{Key Laboratory of Optical Astronomy, National Astronomical Observatories, Chinese Academy of Sciences,Beijing 100012, China}

\author{Shuo Zhang}
\affiliation{Key Laboratory of Optical Astronomy, National Astronomical Observatories, Chinese Academy of Sciences,Beijing 100012, China}
\affiliation{University of Chinese Academy of Sciences, Beijing 100049, China}

\author{Rui Wang}
\affiliation{Key Laboratory of Optical Astronomy, National Astronomical Observatories, Chinese Academy of Sciences,Beijing 100012, China}
\affiliation{University of Chinese Academy of Sciences, Beijing 100049, China}

\author{Li-Li Wang}
\affiliation{Key Laboratory of Optical Astronomy, National Astronomical Observatories, Chinese Academy of Sciences,Beijing 100012, China}
\affiliation{School of Information Management \& Institute for Astronomical Science, Dezhou University, Dezhou 253023, China}
\affiliation{University of Chinese Academy of Sciences, Beijing 100049, China}

\author{Xiao Kong}
\affiliation{Key Laboratory of Optical Astronomy, National Astronomical Observatories, Chinese Academy of Sciences,Beijing 100012, China}
\affiliation{University of Chinese Academy of Sciences, Beijing 100049, China}

\author{Jin-Shu Han}
\affiliation{School of Information Management \& Institute for Astronomical Science, Dezhou University, Dezhou 253023, China}




\begin{abstract}

LAMOST DR5 released more than 200,000 low resolution spectra of early-type stars with S/N$>$50. Searching for metallic-line (Am) stars in such a large database and study of their statistical properties are presented in this paper. Six machine learning algorithms were experimented with using known Am spectra, and both the empirical criteria method(Hou et al. 2015) and the MKCLASS package(Gray et al. 2016) were also investigated. Comparing their performance,  the random forest (RF) algorithm won,  not only because RF has high successful rate but also it can derives and ranks features. Then the RF was applied to the early type stars of DR5, and 15,269 Am candidates were picked out. Manual identification was conducted based on the spectral features derived from the RF algorithm and verified by experts. After manual identification, 9,372 Am stars and 1,131 Ap candidates were compiled into a catalog. Statistical studies were conducted including temperature distribution, space distribution, and infrared photometry. The spectral types of Am stars are mainly between F0 and A4 with a peak around A7, which is similar to previous works. With the Gaia distances, we calculated the vertical height $Z$ from the Galactic plane for each Am star. The distribution of $Z$ suggests that the incidence rate of Am stars shows a descending gradient with increasing $|Z|$. On the other hand, Am stars do not show a noteworthy pattern in the infrared band. As wavelength gets longer, the infrared excess of Am stars decreases, until little or no excess in W1 and W2 bands.

\end{abstract}

\keywords{stars: chemically peculiar --- 
infrared: stars --- catalogs --- surveys --- methods: data analysis --- methods: statistical}


\section{Introduction} \label{sec:intro}
As a class of chemically peculiar (CP) stars, Metallic-line (Am) stars show weaker CaII K lines and enhanced metallic-lines in their spectra than normal A type stars. They were firstly described by \citet{1940ApJ....92..256T}, and were formalized into the MK system by \citet{1948ApJ...107..107R}.  \citet{1970PASP...82..781C}  gave a more detailed definition of Am star describing the nature of itself. These stars, in whose atmosphere, presenting the underabundance of the calcium (or scandium) elements and/or the  overabundance of iron-group elements, are defined as the Am stars. According to the above definition, \citet{1970PASP...82..781C}  divided Am stars into three subgroups, which are stars with both weak CaII K lines and strong metallic lines, stars with only weak CaII K lines, and stars with only strong metallic lines. The common point of the three subgroups of Am stars is that the spectral subtype of CaII K-line is earlier than the metallic-lines. 

Am phenomenon is quite common among  A- and early F-type main sequence stars. The incidence of Am stars exceeds 30\% \citep{1971AJ.....76..896S,1981ApJS...45..437A,2016AJ....151...13G}.  Such a high incidence has attracted the attention of many researchers.  How do Am stars evolve? What characteristics do they exhibit in multiple bands? Are they pulsating? What is their pulsation mechanism?  ...   Due to the limitation of the number of known Am stars, many questions do not have satisfactory answers. E.g, \citet{2017MNRAS.465.2662S} used 864 Am stars to study pulsation and metallicism.  \citet{2017PASP..129d4201A} researched Am stellar evolution on the basis of 462 Am stars.  \citet{2017AJ....153..218C} conducted an infrared photometric study of 426 known Ap and Am stars. \citet{2015MNRAS.448.1378B} investigated light variations of 29 Am stars. However, the numbers of Am stars used in above studies are still too few for their incidence, and the lack of Am stars has become a bottleneck in understanding Am phenomenon.

After the first catalog of the chemically peculiar (CP) stars \citep{1991AAS...89..429R}, \citet{2009AA...498..961R} collected about 4,000 Am stars (or probable) from a large number of literatures and presented another catalogue of CP stars, in which 116 stars have been well studied.  According to an empirical separation curve (ESC) derived from the line index of Ca II K line and 9 groups Fe lines, \citet{2015MNRAS.449.1401H} found 3,537 Am candidates from LAMOST DR1. This is the first search work of Am stars in a large database of low resolution spectra.  However,  Am stars and normal stars can not be distinguished in their marginal region simply by using a separation curve alone.  In the following year, \citet{2016AJ....151...13G} employed  the  MKCLASS program to classify spectra in the LAMOST-Kepler field and totally obtained 1,067 Am stars with hydrogen line types ranging between A4 and F1.The MKCLASS is a spectral classification software that performs the classification through mimicking human-like reasoning, but it was designed for spectra with common type and high quality and sometimes will not succeed on low quality or rare objects. 

A number of  of sky survey projects, such as RAVE, LAMOST, SEGUE, and GAIA-ESO etc., collecting a massive number of stellar spectra which provide us opportunities to search for Am stars. Traditional astronomical research methods such as manual operation and human identification are no longer sufficient.  Many machine learning algorithms have been used in analysis of astronomical data because of their ability to efficiently search and recognize certain type of stars. In this paper, we intend to search for Am stars using machine learning methods.  Compared with various sophisticated  classification algorithms, the RF algorithm wins both in successful rate and efficiency.  In addition, we still need a manual inspection step to guarantee correctness of Am stars obtained by machine learning since the metallic lines in low resolution spectra are very weak and are easily affected by noise. Therefore, we adopt the RF algorithm to design a classifier, and manual examination is used to further check the results.

A key issue is to rule out the contamination by Ap stars, which are also belong to a class of CP stars. Only one or a few elements (including silicon, chromium, strontium, and europium) are extremely enhanced in their stellar atmosphere.  Since some Ap stars also exhibit abundance characteristics of Am stars \citep{1996ApSS.237...77S,2007AstBu..62...62R} to a certain degree, the obtained Am data set may contain a small amount of Ap stars. The largest difference between Am and Ap stars is that Ap stars have intense magnetic fields. However, it is difficult to distinguish between Am and Ap stars in low resolution spectra without spectral features caused by the magnetic effect. In this work, we only can label some spectra with extreme abundance of elements such as silicon, chromium, strontium or europium. Since those elements are also generally enhanced in Am stars \citep{ 2008AA...479..189G, 2008AA...483..891F}, we need follow-up analysis with high resolution spectra to identify whether they are Ap stars or Am stars.   

A large sample of Am stars from a single survey without instrumental or processing differences is useful for statistical study. In this paper, we searched for Am stars in LAMOST DR5 using machine learning methods in conjunction with manual inspection. The paper is organized as follows. In Section 2, we describe the data sets used in this study and data preprocessing steps. In Section 3, we compare various classification methods and show the advantage of RF algorithm in searching for Am stars, and give the result of Am through manually check as well as possible Ap stars. In Section 4, we conduct some physical statistical analysis for the Am stars. Then, a discussion is presented in Section 5. Finally, we summarize this work in Section 6.

\section{Data} \label{sec:data}
\subsection{LAMOST Data} \label{subsec:LAMOST Data}
The Large Sky Area Multi-Object Fiber Spectroscopic Telescope (LAMOST) is a reflecting Schmidt telescope with 4-m effective aperture and 5-degree field of view. 4000 fibers are mounted on the focal plane enable it observe 4000 objects simultaneously. The telescope is dedicated to a spectral survey over the entire available northern sky and is located at the Xinglong Observatory  of Beijing, China \citep{2012RAA....12.1197C,2012RAA....12..723Z,2012RAA....12.1243L,2015RAA....15.1095L}. Compared with SDSS spectroscopic observations, LAMOST survey is more concentrated on the Galactic disk where exists more young stars and is very advantageous to search for Am stars. 

By the end of the first five-year regular survey,  LAMOST has obtained 9,017,844 spectra of stars, galaxies, and QSOs with spectral resolution of $ R\approx 1800 $, wavelength coverage ranging from 370 to 900 nm, and magnitude limitation is about of $r\approx 17.8$ mag for stars. The total number of stellar spectra reaches 8,171,443, making it a gold mine waiting to be exploited. All the above numbers can be found in the LAMOST spectral archive\footnote{\url{http://www.lamost.org}}. 

Before searching for Am stars, we limited the search scope to a certain range through the following conditions:  
  
\begin{enumerate}\label{3 steps}
\item Because spectral type of Am stars are generally A- and early F-type, we limited the search to only A-, F0-, F1-, and F2-type stars in LAMOST DR5, whose spectral types come from the LAMOST 1D pipeline.
\item The spectral features of Am stars mainly appears in blue wavelength band, and their metallic lines are relatively weak and susceptible to noise, thus we only retained spectra with the signal-to-noise ratio of the g-band greater than 50 (S/N$\geqslant $50).
\item  To ensure the accuracy and stability of classification, we eliminated some spectra containing zero flux in the blue spectra. 
\end{enumerate}

More than 10\% objects have been repeatedly observed multiple times by LAMOST. For such targets, we only retain the spectra with the highest signal-to-noise ratio. Finally,  we obtain 193,345 stellar spectra as our searching data set. We need labeled data to form the training and testing data sets, and all the labeled data are also handled using the three operations above except for Evaluation Set II, . 

\subsection{Labeled Data} \label{subsec:labeled data}
In order to train and test the classifier, we collected known samples of Am stars and non-Am samples. We first selected Am samples with confidence greater than 0.5 and all non-Am stars from the work of \citet{2015MNRAS.449.1401H}, and then removed those close to the experience separation curve to ensure the purity of the Am positive and the non-Am negative sample sets.  This yielded 1,805 Am stars as positive samples. For the non-Am stars, further screening was conducted using of MKCLASS software. We  randomly chose the same number of non-Am stars as negative samples, and these positive and negative samples were distributed on average in the training and test sets.  Finally, we obtained 1,806 training samples and 1,804 test samples.

\begin{deluxetable*}{ccll}[h]
\tablecaption{the Labeled Data Sets \label{tab:labeled datasets}}
\tablewidth{0pt}
\tablehead{\colhead{Data Sets} & \colhead{Number of samples} & \colhead{Function}& \colhead{Data Source}} 
\startdata
Training Set & 1,806 & Training classifiers & \citet{2015MNRAS.449.1401H}\\
Test Set & 1,804  & Internal test classifiers & \citet{2015MNRAS.449.1401H}\\
Evaluation Set I & 433  & Extranal evaluating performance of methods & \citet{2016AJ....151...13G}\\
Evaluation Set II & 4  &  Extranal evaluating performance of methods & \citet{2009AA...498..961R} \\ 
\enddata
\tablecomments{ The first column lists the name of the dataset. The second column contains the number of samples in the dataset screened out by various criteria. The third column is the functional description of these datasets in this work. The last column lists the sources of these  datasets.}
\end{deluxetable*}

In addition, we also chose some Am stars  from \citet{2016AJ....151...13G} and \citet{2009AA...498..961R} as evaluation sets in order to evaluate the performance of a variety of methods.   For 1,067 Am stars presented by \citep{2016AJ....151...13G}, we applied the pretreatment process in \ref{3 steps}. Then according to  K-line and metallic-lines spectral subtypes, those samples were classified into classical Am stars and marginal Am stars. The former, K-line type is earlier than the metallic-line type for at least five spectral subtypes. The latter, the difference between the two types is less than five spectral subtypes \citep{2009ssc..book.....G,1978rmsa.book.....M}. We obtained 357 classical Am stars and 76 marginal Am stars as Evaluation Set I. The 116 known Am stars from the catalog of \citet{2009AA...498..961R} were cross-matched  with LAMOST DR5, and only four counterparts were found, which were comprised in Evaluation Set II.
All labeled data sets are summarized in Table \ref{tab:labeled datasets}.

\section{Searching for Am stars} \label{sec:selection}
\subsection{Input Feature Selection} \label{subsec:feature}
According to the characteristics of underabundance of Ca elements and overabundance of Fe group elements in the atmosphere of an Am star, \citet{2015MNRAS.449.1401H} classified Am stars using the empirical separation curve (ESC here after), which is  derived from line indices of K line and 9 groups Fe lines. We used Evaluation Set I  to evaluate the classification ability of the ESC, as shown in Figure 1,  in which there are 357 stars labeled as classical Am stars (green dots in the figure) and 76 stars labeled as marginal Am stars (blue dots). We found that only 345 classical and 52 marginal Ams were judged as Am stars by the ESC (red curve), i.e. the recall through the ESC is 0.966 for classical Am stars and 0.684 for marginal Am stars respectively.

\begin{figure}[ht!]
\center
\includegraphics[width=84mm]{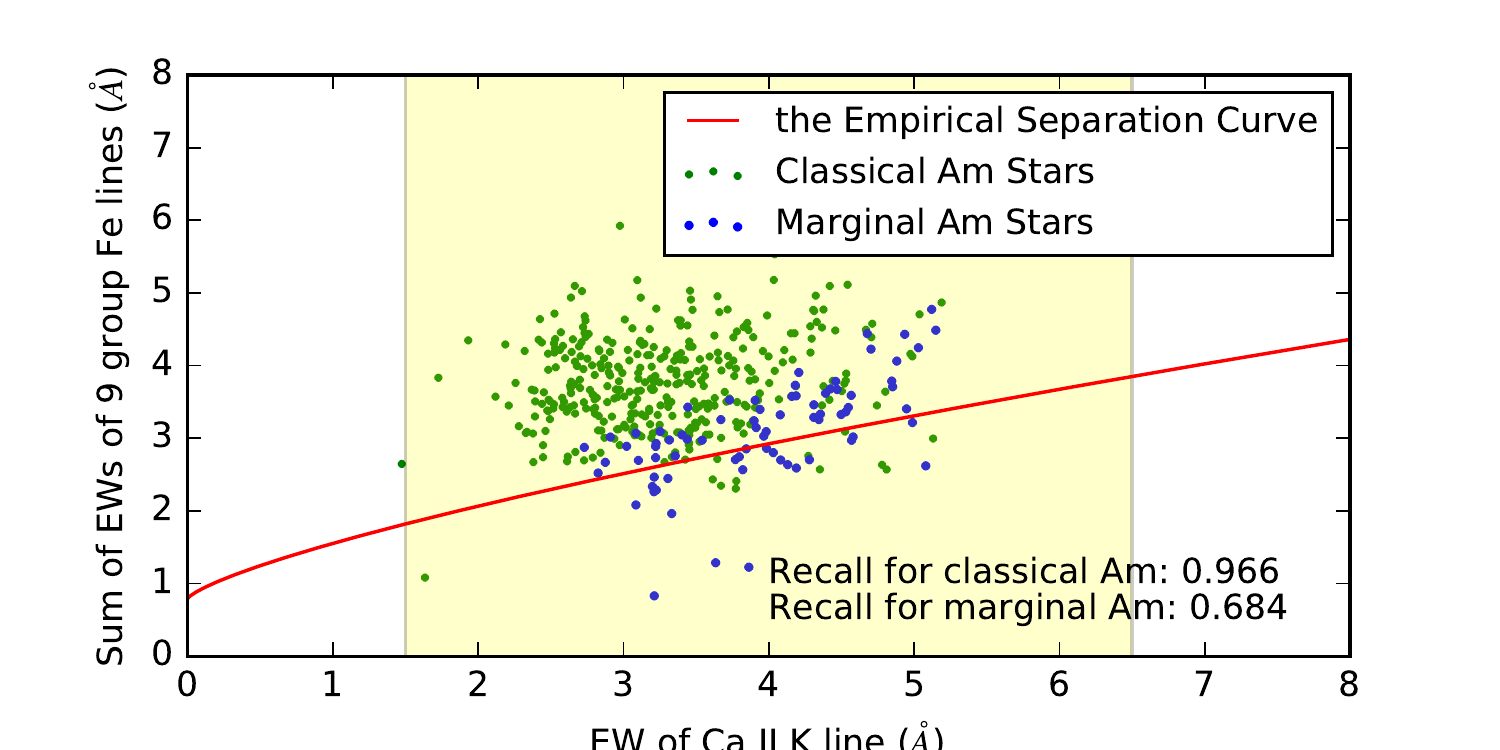}
\caption{Overview of classification performance of the ESC for classical and marginal Am stars. The yellow area indicates the range of  EW(Ca II  K)$\in [1.5, 6.5]$\AA\ . The red curve is the ESC. Those points in the yellow area above the curve are classified as Am stars by \citet{2015MNRAS.449.1401H}. The classification accuracy is 0.966 for classical Am stars and 0.684 for marginal Am stars. \label{fig:hou_data}}
\end{figure}

Obviously, the ESC based on line index method is slightly inadequate for distinguishing marginal Am stars from normal early-type stars. This is mainly because the chemical peculiarity of the marginal Am stars is much weaker than the classical Am stars, and the difference between the marginal Am stars and the normal early-type stars is smaller than that of the classical Am stars.  In addition, some spectral lines, such as Fe lines, are weaker and the batch calculation of those line indices will lead to large errors, which will reduce the recall rate of marginal Am stars.

Based on the above reasons, we decided to use the fluxes of spectra as the input features of classifier. Considering that the spectral line characteristics of Am stars are more densely concentrated in the blue wavelength range than the red, we  choose the normalized fluxes in wavelength range between 3800\AA, and 5600\AA, as the input values of the classifier model.¡±

\subsection{Input Feature Normalization} \label{subsec:continuum}
Before selecting a classification model, we must remove the influence of pseudo-continuum on the classifier. This is the key to successfully distinguish Am stars.  We have improved the fitting technology of pseudo-continuum \citep{2008AJ....136.2022L} by applying the automatic identification operation of strong lines. The details of this procedure are as follows. Step 1, wavelength of all spectra was truncated from  3800\AA\  to 5600 \AA\ . Step 2, a ninth-order polynomial was used to fit each spectrum, and those points that are outside $3\sigma$ away from the fitted function were masked including the strong spectral lines, cosmic rays and sky emissions residual from data reduction. Step 3, a ninth-order polynomial was used  to iteratively fit spectra, where  points  more than $3\sigma$ below the fitted function were rejected. The purpose is to find the approximate upper envelope of each spectrum as its pseudo-continuum. Step 4,  the pseudo-continuum was removed from each spectrum through dividing the observed spectrum by the pseudo-continuum. The intensity of each spectrum was rectified using this method. 

\begin{figure}[ht!]
\center
\includegraphics[width=84mm]{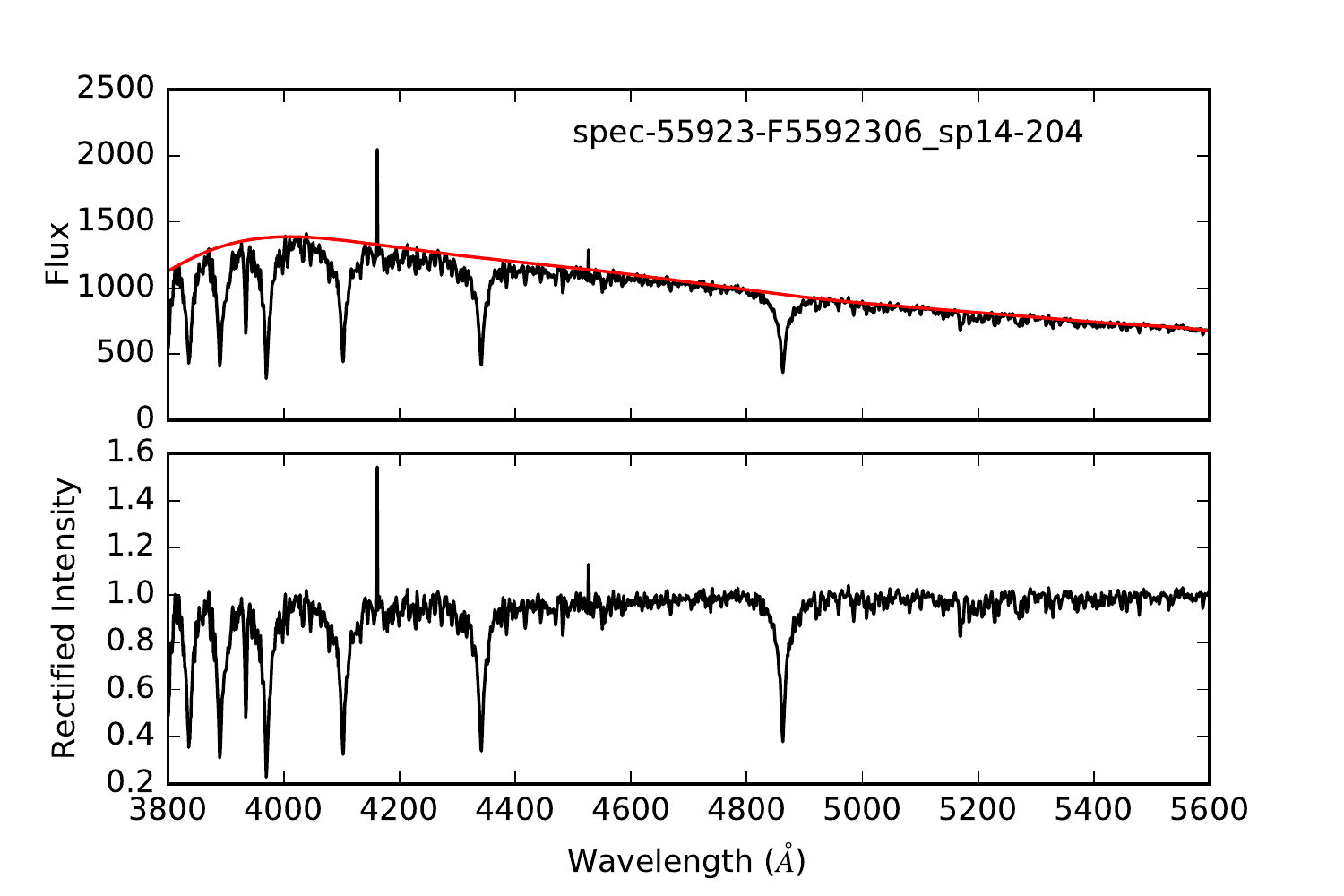}
\caption{An example of rectified flux.  The upper panel shows the truncated original spectrum and the red line shows the fitted contunuum. The red line shows the fitted continuum obtained with the procedure in section 3.2. The bottom panel shows the rectified flux.\label{fig:normalization}. It is note that the emission line in the spectrum would not affect the continuum fitting.} 
\end{figure}

Figure \ref{fig:normalization} shows the results obtained with the improved method.  One can see from Figure \ref{fig:normalization} that the pseudo continuum is well  removed from the spectrum.    

\subsection{Classifier Selection} \label{subsec:classifier}
We selected six sophisticated classification algorithms from scikit-learn web: K-nearest neighbors (KNN), support vector classification (SVC), Gaussian process  (GP), decision tree (DT), RF, and Gaussian naive Bayes (GNB). According to the parameter values recommended by the website, we trained these classifiers on the training set separately and tested their performance on the test set. We then performed two external evaluations for the first three winning classifiers and compared them to the previous  methods of searching for Am stars such as ESC and MKCLASS software, and finally chose the RF algorithm as the classifier in our work.  

\subsubsection{Classifier Evaluation Criteria}
We used precision, accuracy, recall, and F1 score as the criteria to evaluate the classifiers. The four evaluation criteria are defined as follows:
\begin{displaymath}
\begin{split}
&\text{precision} = \frac{\text{TP}}{\text{TP} + \text{FP}} \\
&\text{accuracy} = \frac{\text{TP} + \text{TN}}{\text{TP} + \text{FP} + \text{TN} +\text{FN}}\\
&\text{recall} = \frac{\text{TP}}{\text{TP} + \text{FN}}\\ 
&\text{F1} = 2\times\frac{\text{precision} \times \text{recall}}{\text{precision} + \text{recall}}
\end{split}
\end{displaymath}

Since the test set is composed of labeled samples, it is easy to judge whether the classifier's classification results are correct on the test set. TP is the number of true positive samples that are correctly classified as Am stars by a classifier. FP is the number of false positive samples that are misclassified as Am stars by the classifier. Similarly, TN is the number of true negative samples that are correctly classified as non-Am stars by the classifier, and FN is the number of false negative samples that are misclassified as non-Am stars by the classifier. Precision is the fraction of true positive samples among the set of Am classified by the classifier. Accuracy measures the fraction of samples that are correctly classified in the entire set. Recall measures the fraction of Am that are correctly classified over the total amount of Am. F1 score is the harmonic mean of precision and recall. 

\subsubsection{Internal Testing}
The samples in the training set and the test set come from \citet{2015MNRAS.449.1401H}, among which the positive samples (Am star) were labeled by ESC while the negative samples (non-Am) were labeled by both ESC and MKCLASS software. According to the catalogue of \citet{2015MNRAS.449.1401H}, the positive samples in the training set consisted of 490 (54.3\%) classical Am stars and 413 (45.7\%) marginal Am stars without considering the uncertainty of the spectral subtypes of the K line and metallic lines.  Through the aforementioned classifiers trained on the training set, their classification performance on the test set are listed in Table \ref{tab:scores}. This table is ordered in terms of the F1 scores. Clearly, the first three classifiers (GP, KNN, and RF) show better performance during the internal test. 

\begin{deluxetable*}{ccccccccccccc}[h]
\tablecaption{Performance of the classifiers for test Set\label{tab:scores}}
\tablehead{\colhead{Classifier}  & \multicolumn{4}{c}{Test set}& \colhead{}  & \multicolumn{2}{c}{Classical Am}& \colhead{} &\multicolumn{2}{c}{Marginal Am} & \colhead{} & \colhead{non-Am} \\
\cline{2-5}
\cline{7-8}
\cline{10-11}
\cline{13-13}
\colhead{} &  \colhead{Precision} & \colhead{Accuracy} & \colhead{Recall} & \colhead{F1} & \colhead{}  & \colhead{Accuracy} & \colhead{F1} & \colhead{}  & \colhead{Accuracy} & \colhead{F1}  & \colhead{}  & \colhead{Accuracy}} 
\startdata
GP & 0.998 & 0.996 & 0.994 & 0.996 & & 1.000 & 1.000 & & 0.986 & 0.993 &  &  0.998\\
KNN & 0.998 & 0.995 & 0.992 & 0.995 & & 1.000 & 1.000 & & 0.980 & 0.990 & & 0.998 \\
RF & 0.991 & 0.987 & 0.982 & 0.987 & & 0.998 & 0.999 & & 0.957 & 0.978 & &  0.991 \\
SVC & 0.997 & 0.977 & 0.958 & 0.977 & & 0.971 & 0.985 & & 0.937 & 0.968 & & 0.997 \\
DT & 0.973 & 0.962 & 0.950 & 0.961  & & 0.982 & 0.991 & & 0.900 & 0.947 & &0.973 \\
GNB & 0.893 & 0.911 & 0.935 & 0.913 & & 0.995 & 0.970 & & 0.840 & 0.913 & & 0.888 \\
\enddata
\tablecomments{Here, we only show the accuracy and F1 in classical Am set and marginal Am set, because the recall equals to the accuracy and the precision equals 1 for positive samples. Meanwhile only the accuracy is listed in the non-Am set because the precision, the recall and F1 are all equal 0 for negative samples, }
\end{deluxetable*}

We also divided the test set into three subsets: classical Am, marginal Am,  and non-Am, and tested the performance of the classifiers on them separately. The detailed information is listed in  Table \ref{tab:scores}. As can be seen from the table, the first three classifiers also have good classification performance for the marginal Am stars. 

\subsubsection{External Evaluation I}
The evaluation data set I come from \citet{2016AJ....151...13G}, which consists of 357 classical and 76 marginal Am spectra, and were labeled by MKCLASS software. We tested the classification performance of GP, KNN, RF, and compared them with ESC using the data set. These results are shown in Table \ref{tab:scores1}.  For comparison purposes, the table is ordered in terms of the F1 scores. It can be seen from Table \ref{tab:scores1} that the classification performance of RF is more stable than that of other machine learning algorithms. In addition the classification ability of RF is also more prominent than the ESC method for both classical and marginal Am stars. 

\begin{deluxetable}{ccccccc}[h]
\tablecaption{Performance of classifiers for  Evaluation Set I\label{tab:scores1}}
\tablehead{\colhead{Classifier} & \multicolumn{2}{c}{Classical Am Stars}& \colhead{} &\colhead{} &\multicolumn{2}{c}{Marginal Am Stars} \\
\cline{2-3}
\cline{6-7}
\colhead{} & \colhead{Accuracy} & \colhead{F1} & \colhead{} & \colhead{} & \colhead{Accuracy} &\colhead{F1}} 
\startdata
RF & 0.978 & 0.989 & & & 0.789 & 0.882 \\
GP & 0.972 & 0.986 & & & 0.724 & 0.840 \\
ESC & 0.966 & 0.983 & & & 0.684 & 0.812 \\
KNN & 0.966 & 0.983 & & & 0.671 & 0.803 \\
\enddata
\tablecomments{Here, we only show accuracy scores because accuracy equals recall  and precision equals 1. The table is sorted in descending order according to  F1 scores.}
\end{deluxetable}

\subsubsection{External Evaluation II}

Evaluation Set II were used to compare the RF algorithm with the MKCLASS software. The samples in Evaluation Set II come from the catalog of \citet{2009AA...498..961R}, in which only four counterparts can be found in LAMOST DR5. The RF algorithm and the MKCLASS package were used to classify the four well-studied Am stars, and the results are listed in Table \ref{tab:four sources}. These stars were also recognized as Am in some literatures based on analyzing the abundance of chemical elements and these literatures are also listed in Table \ref{tab:four sources} for reference. The RF algorithm classified these four stars as Am Stars, which is consistent with the results from these literatures.  However, MKCLASS software can only classify the star HD73818 out of the four stars as an Am star. Obviously, the RF classifier is a more suitable tool for searching for Am stars. After all, the MKCLASS software is not a specially developed software for Am stars.

\begin{deluxetable*}{ccccc}[ht!]
\tablecaption{The results of the RF and the MKCLASS with the Evaluation Set II \label{tab:four sources}}
\tablewidth{0pt}
\tablehead{\colhead{HD number} &  \colhead{LAMOST FITS File name} & \colhead{RF} &\colhead{MKCLASS}  & \colhead{Spectral Type of References }} 
\startdata
HD108486  & spec-55959-B5595905\_ sp01-168.fits & Am &  A2 IV-V SrSi &  Am(1,7,8) \\
HD108642 & spec-55959-B5595905\_ sp05-141.fits & Am & A1 IV SrSi &  Am(1,7,8)\\
HD108651 & spec-55959-B5595905\_ sp01-134.fits & Am &A7 mA0 metal weak &Am(6,7,8) \\
HD73818  & spec-57392-KP083141N185915V01\_ sp13-195.fits & Am & kA6hF1mF2  Eu  & Am(2,3,4,5) \\
\enddata
\tablecomments{The first column indicates the identifiers of the stars in the HD catalog. The second column lists the names of FITS files of the LAMOST counterparts. The next two columns show the results of RF algorithm and MKCLASS software respectively. The last column give the literatures that identify the four stars as Am based on element abundance. Articles numbered 1, 2, 3, 4, 5, 6, 7, and 8 corresponds to \citet{2008AA...479..189G}, \citet{2008AA...483..891F}, \citet{2008AA...485..257F}, \citet{2008CoSka..38..123F}, \citet{2007AA...476..911F}, \citet{2006MNRAS.370..819I}, \citet{2004IAUS..224..209M}, and \citet{2000AA...354..216B}, respectively.}
\end{deluxetable*}

\begin{figure*}[ht!]
\gridline{\fig{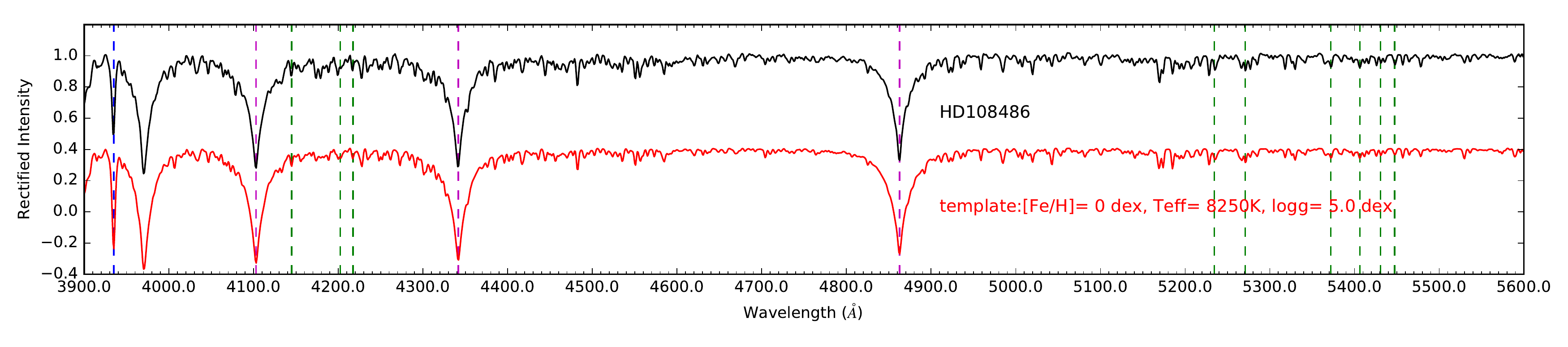}{1\linewidth}{(a)}}
\gridline{\fig{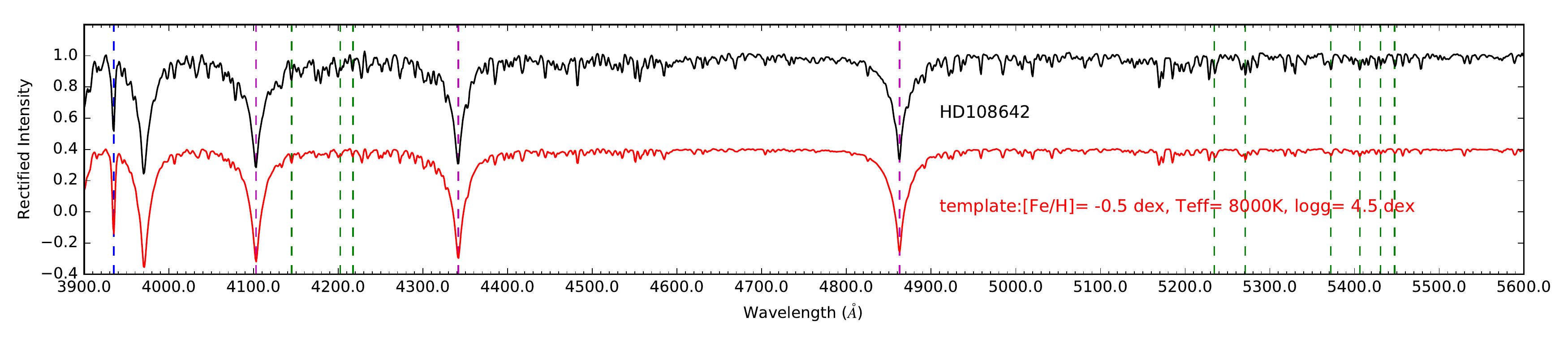}{1\linewidth}{(b)}}
\gridline{\fig{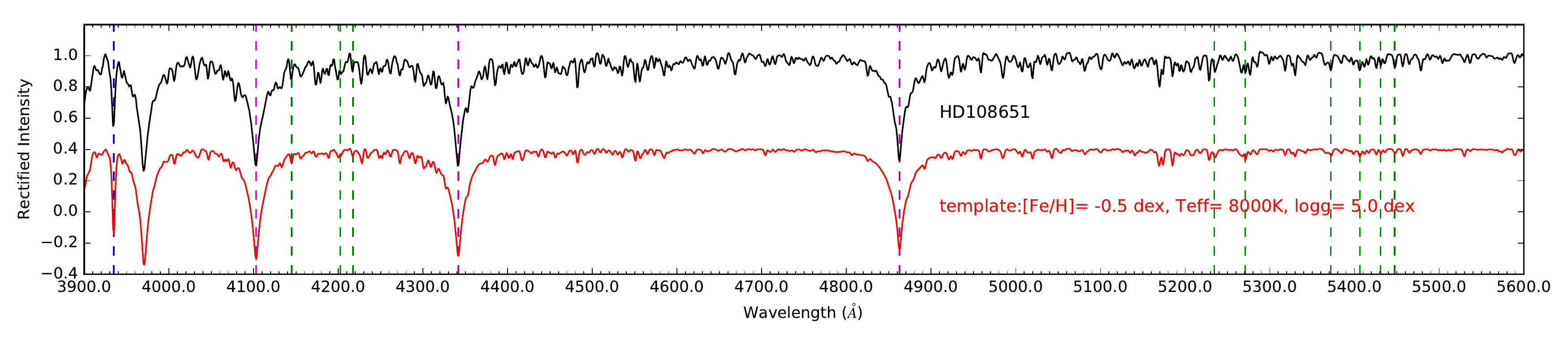}{1\textwidth}{(c)}}
\gridline{\fig{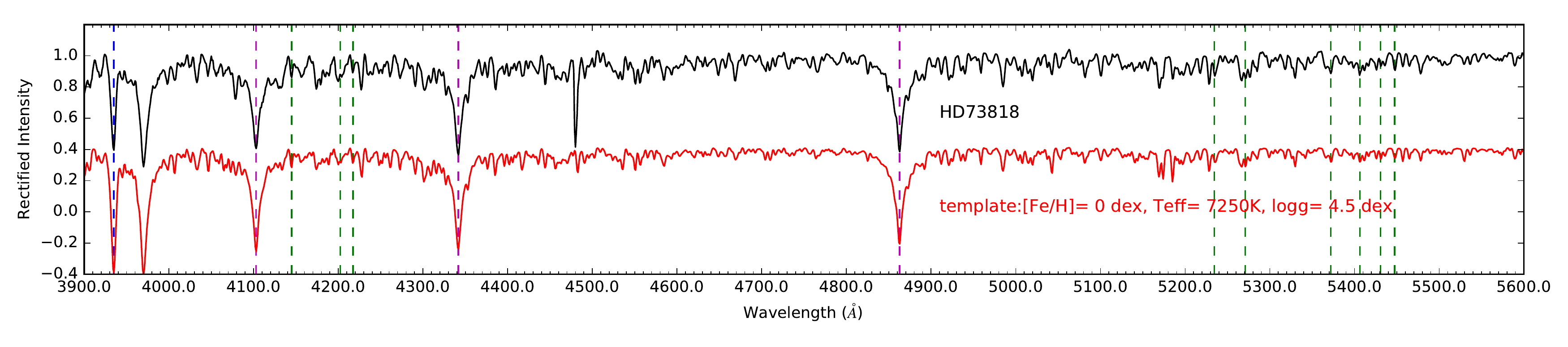}{1\textwidth}{(d)}}
\caption{Comparison between the Am spectra and their best matching templates in Evaluation Set II. The black curves show the Am spectra and the red curve lines are the best matching templates. The name of the Am stars and the atmospheric parameters are also listed next to their corresponding curves in black and red.  The blue, red, and green vertical dashed lines  indicate the positions of the CaII K line, Balmer lines, and some Fe lines from \citet{2015MNRAS.449.1401H}, respectively. All spectra and templates are normalized and the templates have been offset vertically by 0.6 continuum units for clarity. There is an abnormally strong absorption line at around 4480\AA\ of Panel (d), which was caused by bad CCD pixels present in the raw data. The bad pixels were not removed by the data reduction pipeline.}  \label{fig:HD}
\end{figure*}

For visually inspection of the four Am stars, we plot their spectra and corresponding normal stellar templates with the same spectral types given by H lines in Figure \ref{fig:HD}. The best matching Kurucz template \citep{2003IAUS..210P.A20C} for each spectrum was obtained through cross correlation.  The black one in each panel is the normalized Am spectrum while the red one is the best matching template. One can see that the the Balmer lines of the four Am spectra fit well with their best templates, but the strength of the K-lines are weaker than that of their templates, on the other hand the metallic lines show just the opposite. This is in line with the characteristics of  the first subgroup of Am stars with weak K line and strong metallic lines. 

Compared with  ESC, MKCLASS, GP, KNN, SVC, DT, and GNB methods,  the RF algorithm is the best choice to search for Am stars.  After obtaining  Am candidates using RF algorithm, eyeball check was conducted comparing with the best matching templates.

\subsection{RF-Based Classifier} \label{sec:RF}
The RF algorithm a one kind of the bagging algorithm in ensemble learning.  N training samples are randomly selected from the original sample set using the Bootstrapping method with replacement, and K training sets are obtained by K-round extraction. The K training sets are independent of each other, and elements can be duplicated. The K decision tree models are trained on the K training sets, and vote to produce classification results. 

The number of decision trees, K,  is a key parameter in the RF algorithm,  the larger the number of decision trees, the better the classification results, the longer time consumption. After multiple attempts, we used 1800  as the value of the number of decision trees as well as the number of input features. The remaining parameters were set to the default values.

\begin{figure}[ht!]
\center
\includegraphics[width=84mm]{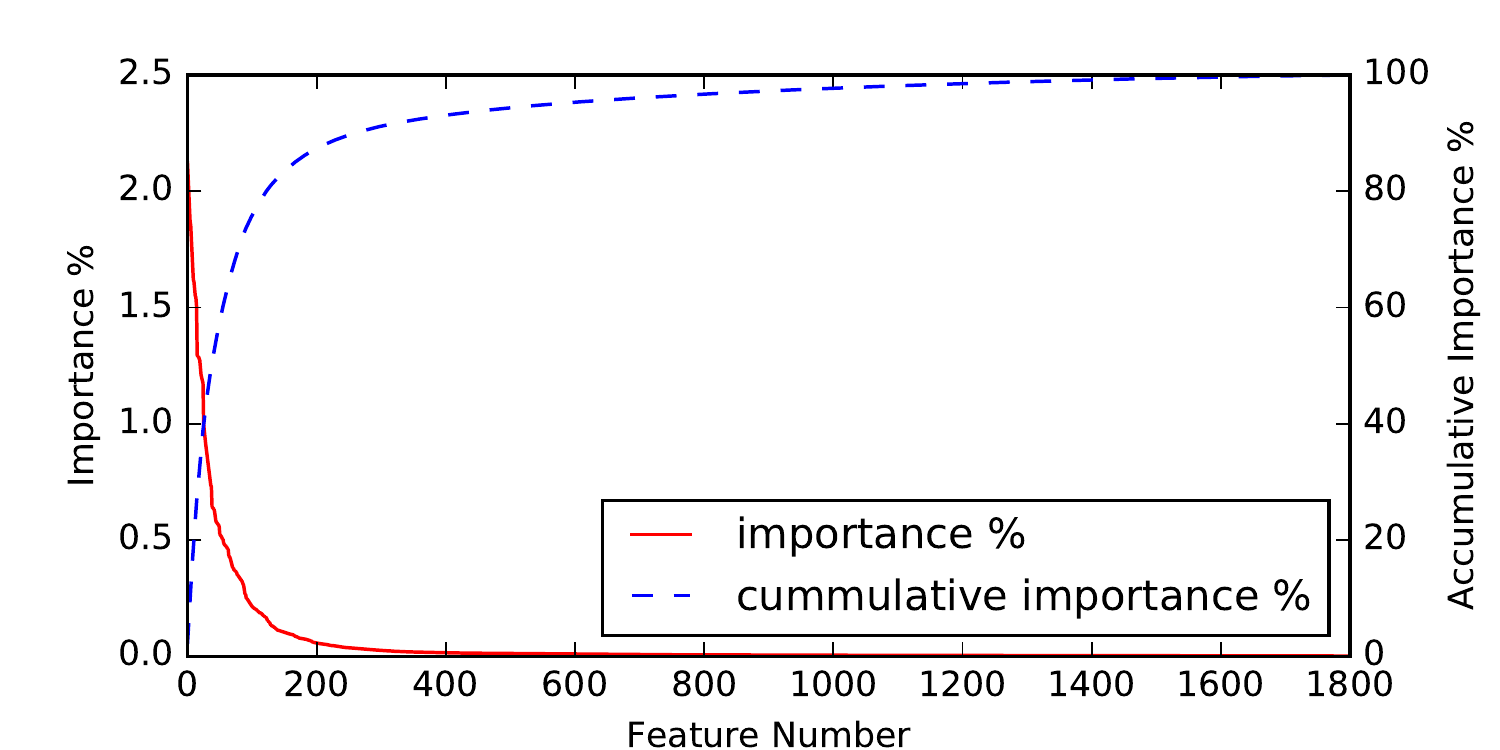}
\caption{Level of importance and accumulated importance as a function of the feature number.  The x-axis is the feature number, and the left y-axis shows the scale of the red solid line, indicating level of the importance of each feature. The right y-axis shows the scale of the blue dashed line, representing the level of the accumulated importance as a function of feature number.} \label{fig:importance}
\end{figure}

\begin{figure*}[ht!]
\center
\includegraphics[width = 190 mm]{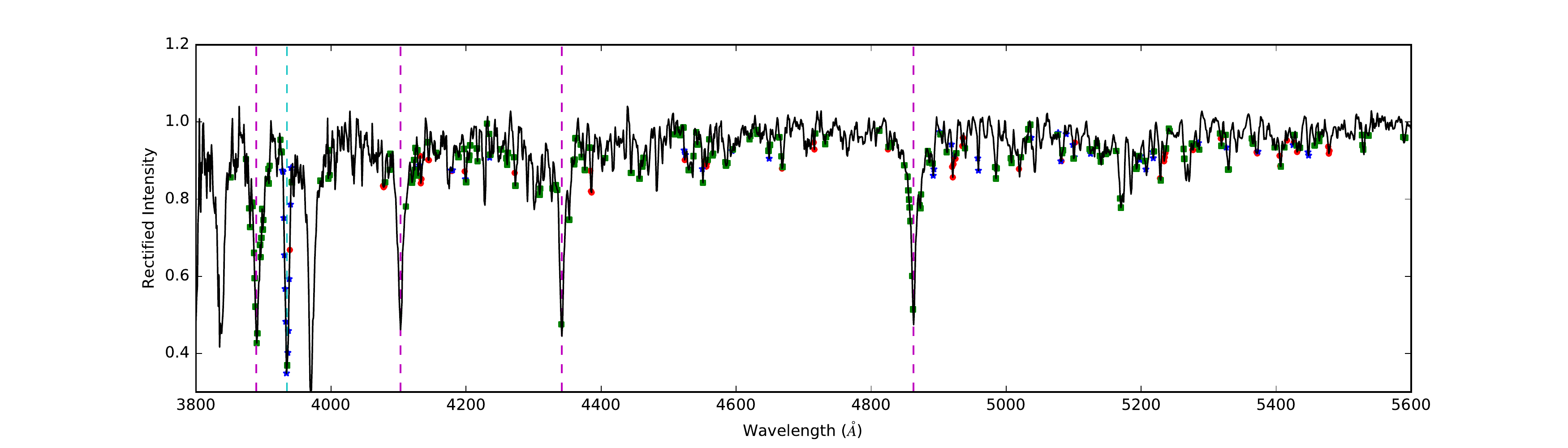}
\caption{Overview of the distribution of first 300 features. The cyan and magenta vertical dashed lines  indicate the locations of the CaII K line and the H lines, respectively. Red dots indicate the first 50 features. Blue stars indicate the next 50 features. Green squares indicate features numbering from 100 to 300.\label{fig:RF_200}}
\end{figure*}

One advantage of the RF algorithm is that it can be used to evaluate the importance of each feature. Figure \ref{fig:importance} shows the importance and accumulative importance of all features.  The importance decreases sharply with the number of features and is almost negligible after number 300.  The first 300 features play important roles in classification, and their accumulated importance reaches 91.2\%. Figure \ref{fig:RF_200} shows the distribution of the first 300 features in a spectrum.   Those features basically fall on the absorption lines of CaII K, H, and transition metal elements, which are considered to be very important elements distinguishing Am stars. 

We identified the spectral lines where the first 50 feature points are located by consulting the line table of the theoretical model of Moore and other early-type stars literatures \citep{2017AA...604L...9P,1994MNRAS.271..355A,1986AAS...64..477C,1973ApJS...25..277S}. The details are listed in Table \ref{tab:feature_50}. We only listed the main elements contained in spectral lines since metal lines in low-resolution spectra are mostly blended lines. It should be noticed that feature numbers and importance are not absolute. Different RF algorithm classifiers will produce different results because the data used by RFs in constructing each decision tree are randomly selected from the training set. Fortunately, their ranking and importance do not change much for most of the important features.  It should also be noted that the wavelengths of the features are all in vacuum, because spectra of LAMOST are all converted to vacuum wavelength, and you can find the relevant  keyword ``VACUUM" in the FITS header of the spectra.

\begin{deluxetable*}{lclclcl}[ht!]
\tablecaption{Identification of relevant spectral lines based on the location of the first 50 feature points. \label{tab:feature_50}}
\tablehead{\colhead{ID} &\colhead{} & \colhead{Name} &\colhead{} & \colhead{Valuum Wavelength}&\colhead{} & \colhead{Importance}} 
\startdata
1, 11, 24 &  & BaII, FeI  & & 4555.292, 4557.394 & &2.121, 1.61, 1.187 \\
2, 3, 18, 26  &  & FeI &  & 4921.871 & & 2.050, 2.021, 1.289, 0.997\\
4, 16, 36 &  & FeI, NiI &  & 5478.086, 5478.432 & & 1.882, 1.295, 0.741\\
5, 9, 38 &  & FeI &  & 4384.766 & & 1.87, 1.701, 0.729\\
6, 10 &  & FeII, CrI, CoI &  & 5277.457, 5277.526, 5277.629 & & 1.853, 1.620\\
7, 17, 19 &  & SrII &  & 4078.849 & & 1.797, 1.289,  1.279\\
8, 33, 43 &  & FeI &  & 5234.395 & & 1.77, 0.828, 0.621\\
12, 13 &  & FeI, NiI &  & 4715.671, 4715.720 & & 1.579, 1.551\\
14, 46 &  & FeI, NiI &  & 5372.982, 5372.82 & & 1.543, 0.576\\
15 &  & FeI, FeII &  & 4523.775, 4523.885 & & 1.506\\
20, 23, 49 &  & FeI &  & 4133.207 & & 1.273, 1.197, 0.564\\
21, 22 &  & FeI, BaII &  & 4935.391, 4935.456 & & 1.232, 1.21\\
25 &  & FeI &  & 4178.763 & & 1.174\\
27 &  & FeI &  & 5416.703 & & 0.964\\
28, 40 &  & FeI &  & 4145.021 & & 0.949, 0.638\\
29 &  & FeI, NiI &  & 5100.113, 5101.343 & & 0.916\\
30, 35 &  & FeI &  & 5407.276 & & 0.892, 0.778\\
31 &  & FeI &  & 5425.576 & & 0.871\\
32 &  & FeI &  & 4669.36 & & 0.855\\
34 &  & FeII &  & 5318.086 & & 0.800\\
37 &  & FeI &  & 5228.634 & & 0.736\\
39 &  & MnI, FeI &  & 4824.844, 4825.473 & & 0.642\\
41 &  & FeII &  & 4925.288 & & 0.636\\
42 &  & FeII &  & 5019.834 & & 0.633\\
44 &  & CaII K &  & 3934.767 & & 0.592\\
45 &  & FeI &  & 4272.952 & & 0.580\\
47 &  & FeI, NiI &  & 5080.631, 5081.94 & & 0.570\\
48 &  & FeI &  & 5431.203 & & 0.566\\
50 &  & FeI &  & 4199.4 & & 0.560\\
\enddata
\tablecomments{The first column lists the feature ID. In order to make the table more concise, features that fall on the same absorption line are placed in the same entry. The second column shows the name of the line in which feature points are located.  The third column lists the vacuum wavelength corresponding to each spectral line. The fourth column shows the importance of the corresponding feature determined with the RF algorithm.}
\end{deluxetable*}

\subsection{Manual Inspection} \label{subsec:inspection}

Three reasons require manual inspections of the Am candidates obtained using RF method: First, the intensities of metal lines in Am spectra are very weak and are easily masked by noise, which would lead to errors in the results. Second, although the spectra were rectified, the residual continua still could affect the classification. Third, the precision of RF algorithm is 0.991, means there are still a small fraction of stars might be wrongly recognized. 

The specific process of manually inspection is to compare the spectral lines of those candidates with their best matching synthetic template, a set of quantitative standards is as follows:

\begin{align*}
\begin{cases}
\quad EW(K_{spe}) < EW(K_{mod})  \nonumber\\
\quad \sum{EW(M_{spe})} = \sum{EW(M_{mod})} \\
\end{cases} \\
or \\ 
\begin{cases}
\quad EW(K_{spe}) = EW(K_{mod})  \nonumber\\
\quad \sum{EW(M_{spe})} > \sum{EW(M_{mod})} \\
\end{cases} \\
or \\ 
\begin{cases}
\quad EW(K_{spe}) < EW(K_{mod})  \nonumber\\
\quad \sum{EW(M_{spe})} > \sum{EW(M_{mod})} \\
\end{cases} 
\end{align*}

where, $K_{spe} $ and $K_{mod} $  are the CaII K lines of a spectrum and its matching template, respectively. $M_{spe} $ and $M_{mod} $  are metallic lines of a spectrum and its matching template, respectively.  $ EW(\cdot)$ is the equivalent width of a spectral line. 

\begin{figure*}[ht!]
\center
\includegraphics[width=190mm]{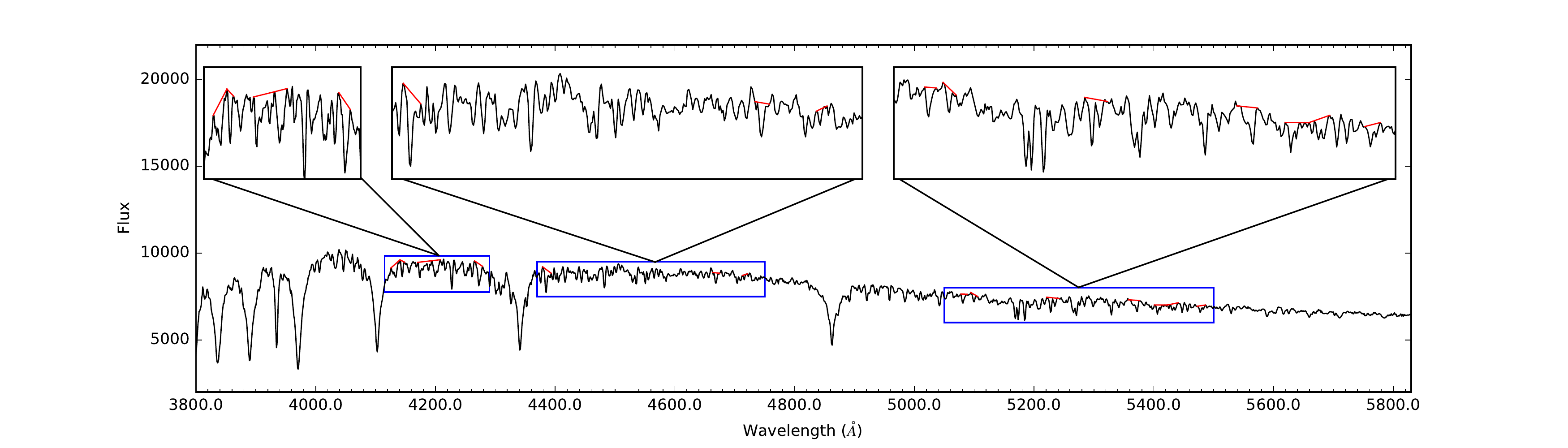}
\caption{ Local continuum spectrum of 15 Fe-group lines. \label{fig:15_Fe}}
\end{figure*}

In this work, we adopted the same EW definition of the CaII K line as \citet{2015RAA....15.1137L}, the line is in the window [3927.7\AA-3939.7\AA], and the blue and red sideband are in [3903\AA-3923\AA], [4000\AA-4020\AA] respectively. For metallic lines, conventional EW calculation is not suitable because the Fe absorption in A-type stars is generally weak and too narrow to give wavelength bands that EW needs.  So we had to use the method proposed by \citet{2015MNRAS.449.1401H} to calculate their equivalent width. We selected part of FeI lines listed in Table \ref{tab:feature_50} by eliminating several FeI lines blended with ionized elements. We merged adjacent Fe lines into 15 Fe-group lines and listed the left ends and the right ends of these groups in Table \ref{tab:Fe_wavelength}. To calculate the EW, side bands for blue and red are defined in [Left\_End-5\AA, Left\_End],  [Right\_End, Right\_End+5\AA] respectively. We limited the sidebands to 5\AA\ to get the best local pseudo continuum for each of the 15 Fe-group lines avoiding affection by other lines.  Figure \ref{fig:15_Fe} shows an example of a local pseudo continuum of 15 Fe-group lines. 

\begin{deluxetable}{ccc}[h]
\tablecaption{Wavelength ranges of 15 Fe-group lines.\label{tab:Fe_wavelength}}
\tablehead{\colhead{Number} & \multicolumn{2}{c}{Center Band}\\
\cline{2-3}
\colhead{} & \colhead{Left End (\AA)} & \colhead{Right End (\AA)}} 
\startdata
1 & 4128 & 4138 \\
2 & 4142 & 4146 \\
3 & 4171 & 4192 \\
4 & 4194 & 4207 \\
5 & 4268 & 4278 \\
6 & 4383 & 4390 \\
7 & 4668 & 4671 \\
8 & 4714 & 4718 \\
9 & 5079 & 5087 \\
10 & 5097 & 5104 \\
11 & 5225 & 5240 \\
12 & 5362 & 5375 \\
13 & 5405 & 5420 \\
14 & 5424 & 5439 \\
15 & 5477 & 5485 \\
\enddata
\tablecomments{All values are in vacuum wavelengths.}
\end{deluxetable}

Although there are some uncertainty in calculating the equivalent width in batches, such as noise interference in Fe lines or mixing of other metal lines in the CaII K line wing, the line ratio of the CaII K to $H\varepsilon $ or $H\delta $ can be verified as important criteria during manual inspection. 

\subsection{Labelling Ap stars} \label{subsec:Ap_star}
There is some contamination by Ap stars to the obtained Am candidates because of similar spectral features. Most of the Ap stars are actually B-type stars, while a small portion of  Ap stars are also  found in A- and F-type stars. Among those cooler Ap stars,  some also exhibit the characteristics of the overabundance of Fe elements and the underabundance of Ca elements. Therefore, a small amount of Ap stars will be mixed with Am stars.

According to the definition of Ap star, we thought of objects whose Sr, Cr, Eu, or Si element  are extremely abundant as Ap candidates.  In general, the abundance of  Sr, Cr, Eu, and Si in Am stars rarely exceeds 2.0 relative to the abundance of the sun \citep{2015MNRAS.451..184C,2014MNRAS.441.1669C,1996ApSS.237...77S,1987ApJS...65..137L}. Therefore stars with Sr, Cr, Eu, or Si element abundance exceeding 2.0 are likely to be Ap stars. The detailed method of finding Ap candidates is as follows:  Firstly, according to the  prominent spectral lines in the blue-violet spectra of Ap stars \citep{2009ssc..book.....G}, and excluding some lines with nearby FeI lines or other line interference, we  selected the 4077\AA\,  line as a reference line, which is a blending line of SrII, CrII, and SiII in LAMOST spectra. Secondly, we generated corresponding synthetic templates by setting the values of [Sr/H], [Cr/H], [Eu/H], and [Si/H] as 2.0 in the spectrum generator SPECTRUM. Stellar parameters that Kurucz model required, such as $T_{\rm eff}$, $log g$ and $[Fe/H]$, were taken from the corresponding Am models which have normal abundance of above mentioned heavy elements.. Finally, we use the method of \citet{2015MNRAS.449.1401H} to calculate the EW(tem) and EW(obs) of the 4077\AA\ blend line for both the templates and observed Am spectra, and marked those objects as Ap candidates if their EW(obs)$>$ EW(tem).

\subsection{Result} \label{subsec:result}

For 193.345  spectra of the searching data set described in Section \ref{subsec:LAMOST Data}, we simply fitted with Kurucz templates in the wavelength range of [3900\AA, 5600\AA] to obtain $T_{\rm eff}$ and $log g$. With the stellar parameters, we can further constrain the searching data set with 6500 K$< T_{\rm eff} \leqslant $11000 K and  $log g \geqslant  4.0$ dex, since the Am phenomenon often occurs in A- and early F-type main-sequence stars. By this constraint,  98,202 spectra were retained, and then the RF classifier was applied to identify 15,269 Am candidates, for which we carried out manual inspections. After these inspections we discarded 4,766 spectra, among which 1,338 candidates (28\%) do not meet the reference criteria of Section \ref{subsec:inspection}, 2,585 spectra (54\%) cannot be recognized by human eyes because of their small peculiarity, and 843 spectra (18\%) are of bad spectral quality and not sufficient to identify the Fe line. In addition, using the method described in Section \ref{subsec:Ap_star}, we found 1,131 objects with extreme overabundance of Sr, Cr, or Si elements and labeled them as Ap candidates in the catalog. Whether or not they have the nature of Ap needs to be identified by subsequent analysis of their magnetic field strength. In total the catalog has 10,503 entries including 9,372 Am stars and 1,131 Ap candidates. In the statistical analysis section below, we excluded them from Am stars.

For each Am star in the catalog, we also determined three different spectral subtypes of its K-line, H-lines, and metallic lines using the template matching method. The band used to match the spectral subtype of metallic-lines is the combined band of [4140\AA, 4300\AA], [4410\AA, 4600\AA], and  [4900\AA, 5400\AA]. The matching band for H-lines is a combination of the $H\beta $, $H\gamma $, and $H\delta $  bands.It is worth noting that the spectral subtype obtained by matching with templates in the specific wavelength ranges alone are not completely accurate, and the spectral subtypes of some Am stars do not conform to the common characteristics of Am stars, i.e. the K-line spectral subtype is earlier than the metallic-line spectral subtype. This is why we can not use this criterion for Am search directly. 

The complete catalog of identified Am stars can be downloaded http://paperdata.china-vo.org/Qinli/2018/
dr5\_Am.csv, and the example catalogue is presented in Appendix A. 

\section{Statistical Analysis} \label{sec:analysis}

\subsection{Effective Temperature Distribution} \label{subsec:incidence}

\begin{figure}[ht!]
\center
\includegraphics[width=85mm]{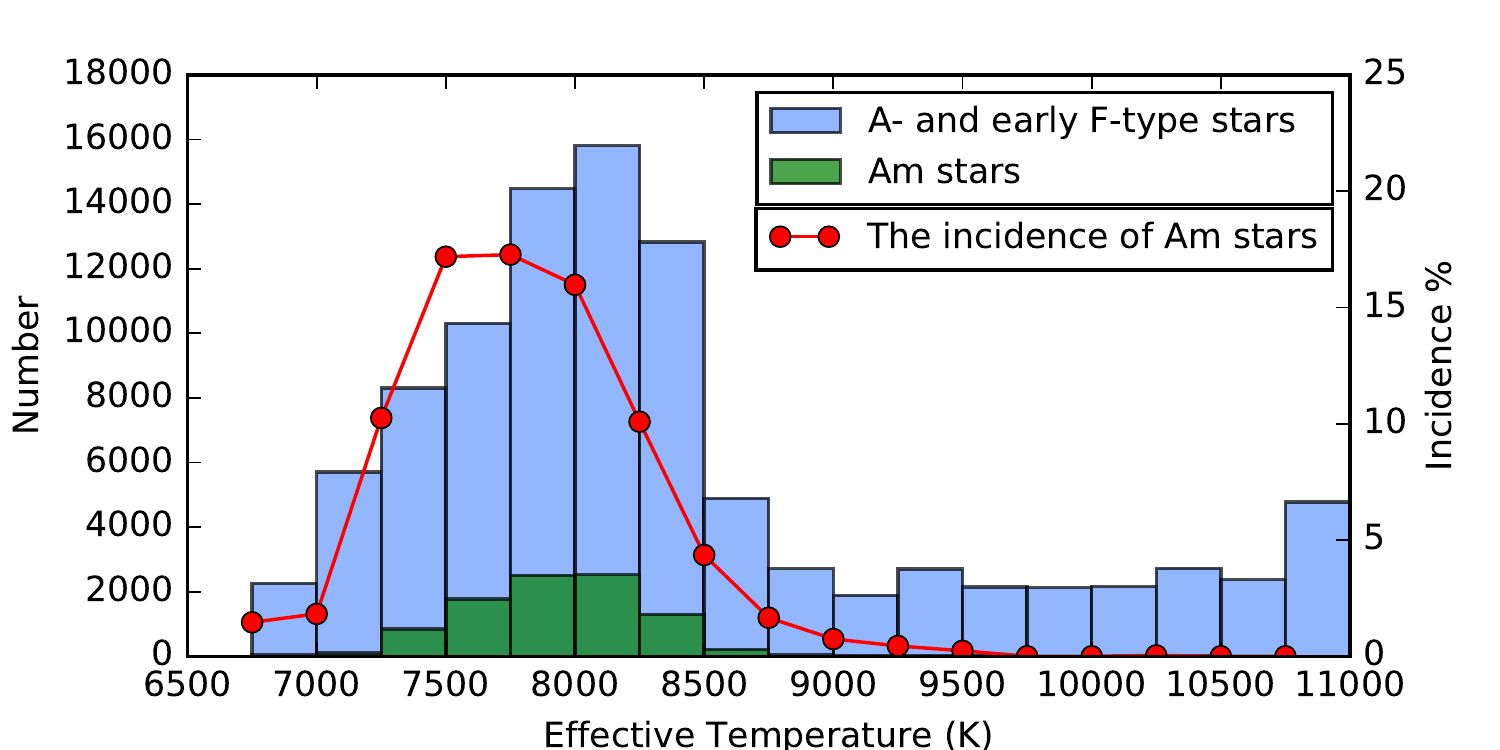}
\caption{Distribution and Incidence of Am stars as a function of effective temperature. The y-axis on the left indicates the distribution of Am stars and the y-axis on the right is the frequency of occurrence of Am stars. The red points is the incidence of Am stars in each bin, they are displayed to the left of each bin.  \label{fig:incidence}}
\end{figure}

We analyzed the effective temperature distribution for these Am stellar samples. Figure \ref{fig:incidence} shows that the distribution of Am stars and the incidence of Am stars in different effective temperature bins.  As can be seen from Figure  \ref{fig:incidence}, the results are  consistent with  the conclusion presented by \citet{2017MNRAS.465.2662S}, namely  that the temperature of Am stars is mostly distributed between 7250 K (F0) and 8250 K (A4), peaking  near 7750 K. Due to our strict screening, the fraction of Am stars  to the total A- and early F-type stars is smaller than the values reported in previous studies \citep{1971AJ.....76..896S,1981ApJS...45..437A,2016AJ....151...13G}. The incidence of Am stars we give can be used as the lower limit.

\subsection{Space Distribution} \label{subsec:distribution}
\begin{figure}[ht!]
\center
\includegraphics[width = 95 mm]{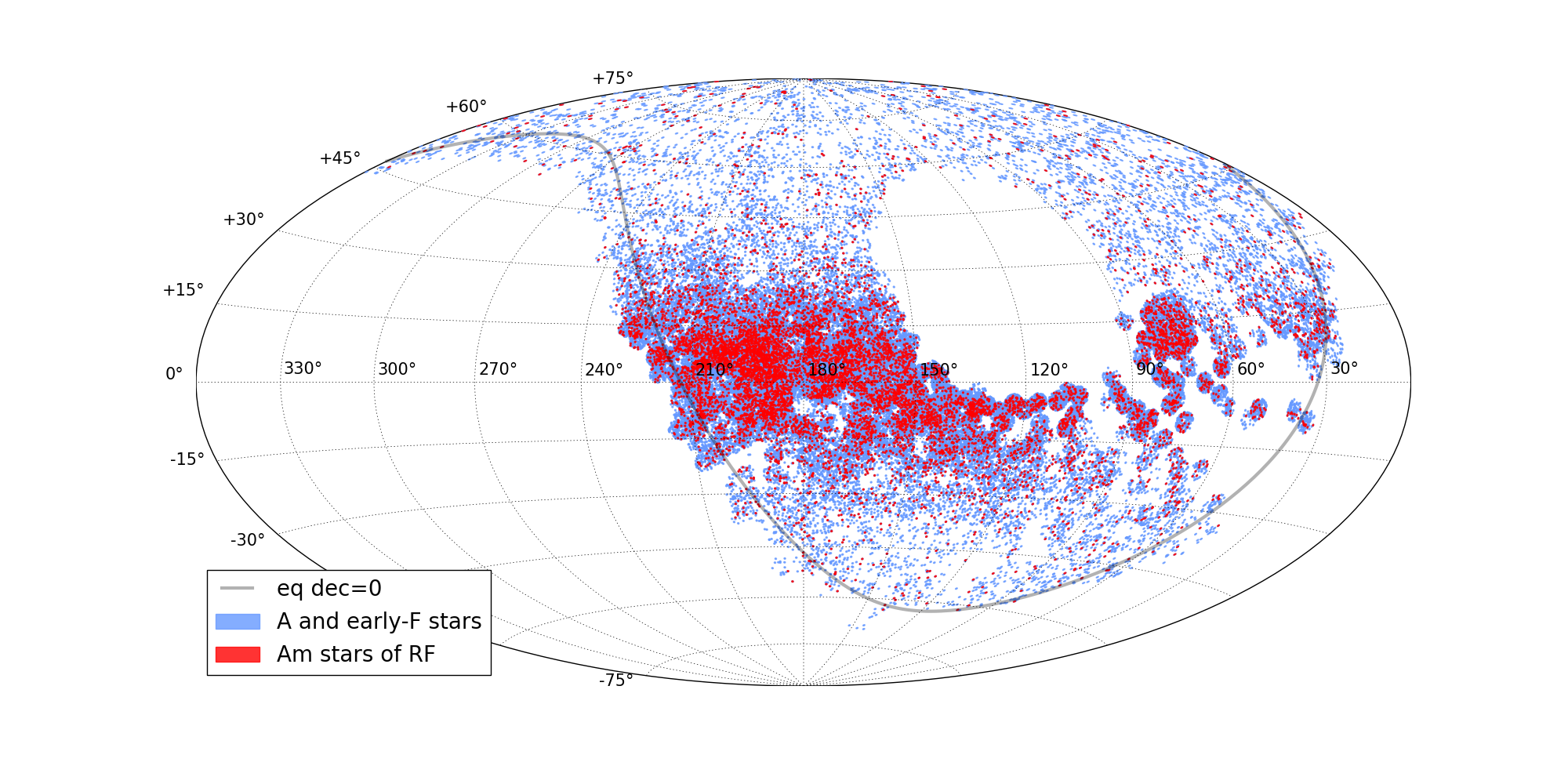}
\caption{Spatial distribution of Am stars on the Galactic coordinate plane. The blue and the red dots correspond to all early type stars and Am stars.\label{fig:space}} 
\end{figure}

The spatial distribution of Am stars in the Galactic coordinate plane is plotted in Figure \ref{fig:space}. The blue points indicate all A- and early F-type stars, and the red points are for Am stars. As shown in Figure \ref{fig:space}, the number of Am stars on the Galactic disk is significantly higher than that in other regions, because most of the observations of LAMOST  are concentrated on the Galactic disk, especially in the anti-center direction.  The LAMOST Spectroscopic Survey of the Galactic  Anti-center (LSS-GAC), which covers  Galactic longitudes $ 150\degr \leq \ell \leq 210 $ and latitudes $|b| \leq 30\degr $, is a unique component of LAMOST Experiment for Galactic Understanding and Exploration (LEGUE) spectroscopic survey \citep{2015RAA....15.1095L}. 

\begin{figure}[ht!]
\center
\includegraphics[width=84mm]{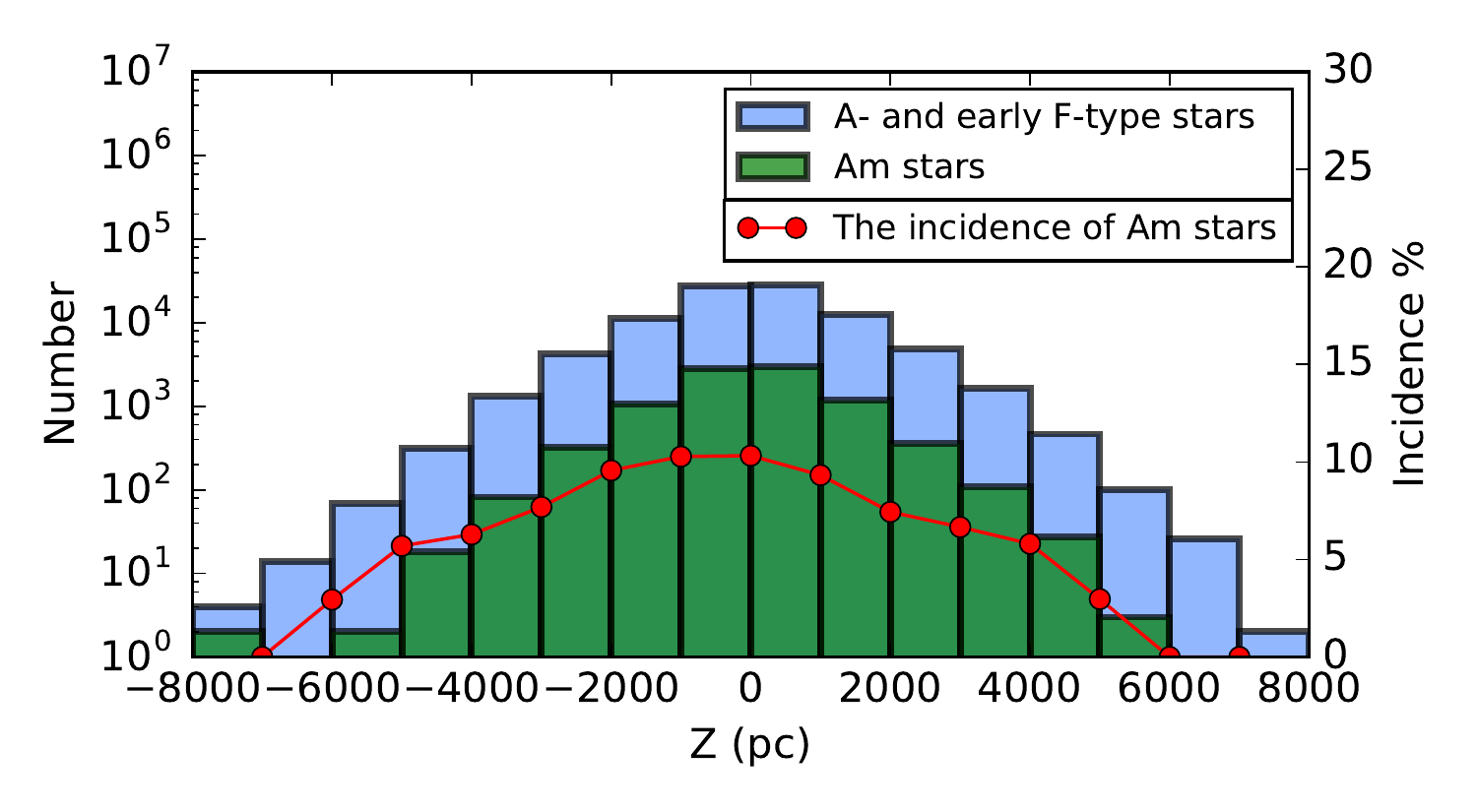}
\caption{Distribution and incidence of Am stars as a function of vertical distance from the Galactic plane Z. The red points on the left to each bin are the incidence of Am stars, and we did not compute the incidence for the bins in which there are  less than 10 early type stars because of no statistical meaning.  \label{fig:space_z}}
\end{figure}

In order to further understand the spatial distribution of Am stars, we analyze the  frequency of occurrence of Am stars  as a function of the vertical distance from the Galactic plane (Z). We obtained the parallax($\omega$)  of most spectra by cross-matching with the Gaia DR2. For spectra with parallax $\geqslant0$, we then cross-matched with the catalog in \citet{2018AJ....156...58B} and got their estimated distances. Eventually, we obtained the  distances of 92,870 early-type stars and 8,951 Am stars. The vertical distance Z for each star can be calculated with the following formula:

\begin{displaymath}
Z = r \times \sin(b)
\end{displaymath}

where $b $ is the Galactic latitude, and $r $ is the estimated distance. In Figure \ref{fig:space_z}, the blue and green histograms show the distribution of early-type stars and Am stars along the vertical distance $|\text{Z}|$, respectively. In each bin, we calculated the incidence of Am stars and represented them with red points. Figure \ref{fig:space_z} suggests that the incidence of Am stars increases as $|\text{Z}|$ decreases.

\subsection{Infrared Photometry} \label{subsec:infrared excesses}
We also performed infrared photometric analysis on these Am stars.  First, we cross-matched them with 2MASS and WISE catalogs with a matching radius of 3.0 arcsec, and  obtained the corresponding magnitudes of the J, H, K, W1, W2, W3, and W4 bands. Because the WISE satellite has low angular resolutions of 6.1, 6.4, 6.5, and 12.0 arcsec in the W1, W2, W3, and W4 bands, we often found that one 2MASS counterpart is a different object to the WISE counterpart. In order to avoid this case, multiple WISE sources within a search radius of 10.0 arcsec were eliminated. In order to improve the accuracy of the result, the photometric errors were limited to less than 0.1 mag in all three 2MASS bands \citep{2006AJ....131.1163S} and 0.05 mag in the W1 and W2 bands \citep{2010AJ....140.1868W}. Then, the color excess of all data in the color $(a-b)$ are estimated using the formula: $E(a-b) = R(a-b) \times E(B-V) $, where $E(B-V)$ values are given in  \citet{2011ApJ...737..103S} and the $R(a-b) $ values are reddening coefficients of the color $(a-b) $ from \citet{2013MNRAS.430.2188Y}. We calculated color excess and dust reddening of 7,799 Am stars for the (J-H), (H-K), and (W1-W2) colors.  Finally, our conclusions are shown in Table \ref{tab:infrared_excess}.  One can see a very clear downward trend in the incidence of infrared excess from near-infrared to mid-infrared, and the incidence reduces to 0.15\% in the W1 - W2 region. 

This is in contradiction with the conclusion about Am stars from \citet{2017AJ....153..218C}. They found that over half of Am stars have clear infrared excess ($(W1-W2) > 0.1 $) in the W1 - W2 region and have no or little infrared excess in the remaining regions, including J, H, K, and IRAS regions. We checked the data set from \citet{2017AJ....153..218C} and found that they do not restrict the photometric precision to $W1_{\rm error}<$ 0.05 mag and $W2_{\rm error}< $0.05 mag. When we add this constraint, there are only 3 sources in Chen's Am dataset with infrared excess. Thus, we statistically conclude that Am stars have no infrared excess in the W1 - W2 region.

\begin{deluxetable}{Ccc}[h]
\tablecaption{ Results of infrared excess analysis \label{tab:infrared_excess}}
\tablehead{\colhead{Criterion} & \colhead{The Number of Am Stars} & \colhead{Incidence\tablenotemark}} 
\startdata
$J-H>0.1 $ & 1652 & 21.68\% \\
$H-K>0.1 $ & 163 & 2.09\% \\
$W1-W2>0.1 $ & 12 & 0.15\% \\
\enddata
\end{deluxetable}

\section{Discussion}
Generally, [Fe/H] is often used to relatively represent the metallicity of a star. However, compared to normal stars, the atmosphere of an Am star is Fe enriched and Ca deficient. The metallicity of an Am star obtained with conventional methods may be larger than the true value. Taking into account that Am stars comprise a significant fraction of early-type stars, researchers should take much care about the metallicity of Am stars given by pipelines for spectral surveys especially for statistical study.

\begin{figure}[ht!]
\center
\includegraphics[width=84mm]{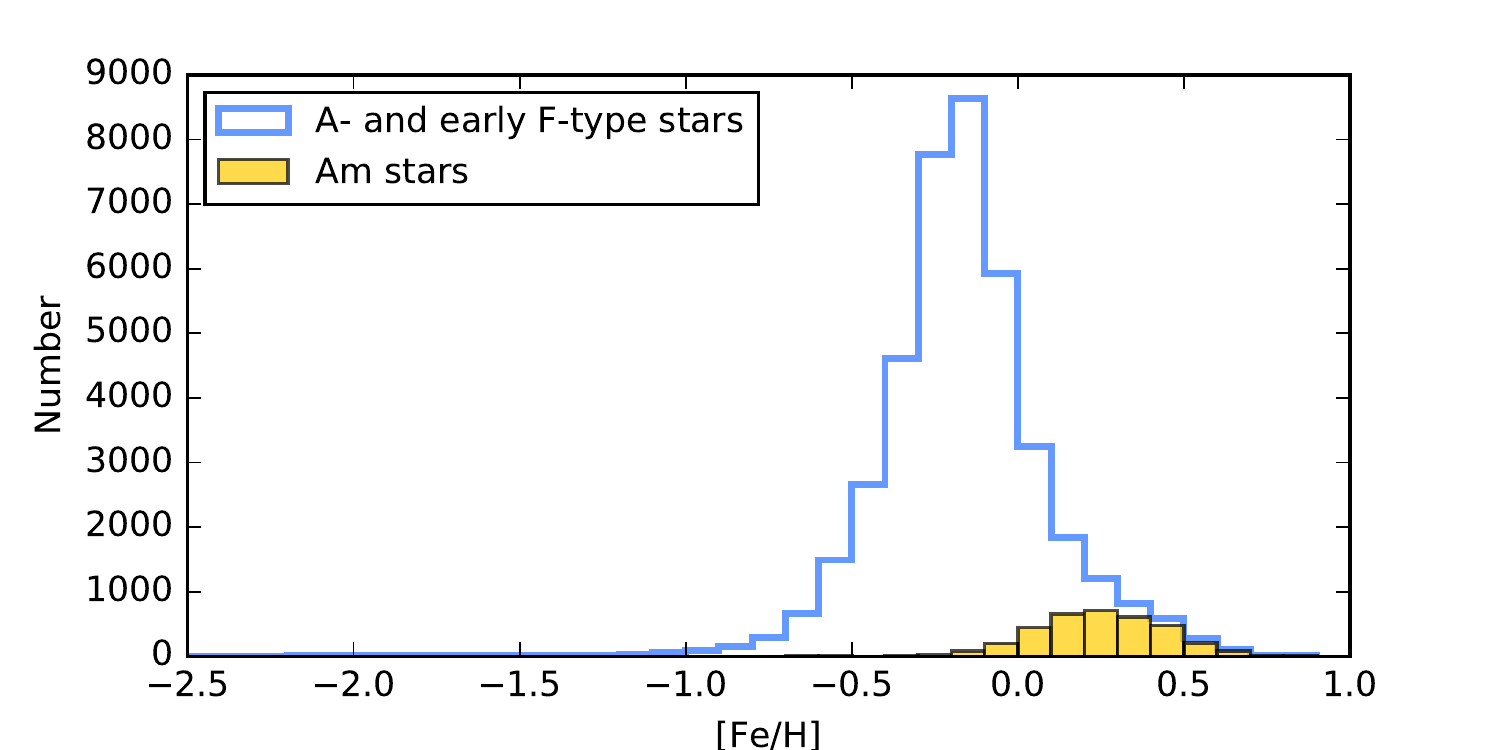}
\caption{Distribution of metal abundance in early-type stars and Am stars.  The metallicity parameters were taken from the LAMOST catalog.  \label{fig:metal}}
\end{figure}

In order to understand the degree of  metallicity overestimation in Am stars,  we analyzed the metallicity distribution of Am stars with LAMOST atmospheric parameters. The metallicity given by LAMOST pipeline \citep{2015RAA....15.1095L} causes Am stars as a whole to be biased toward metal enrichment relative to normal early-type stars. A detailed statistical metallicity distribution ([Fe/H]) is shown in Figure \ref{fig:metal}. The blue histogram shows the distribution of [Fe/H] for all A- and early F-type stars with the LAMOST atmosphere parameters. The yellow correspond to the Am stars with LAMOST atmosphere parameters.  One can note that the right region of  the Figure \ref{fig:metal} is dominated by Am stars. The conclusion that most metal-rich stars are Am stars is obviously unreasonable. 

\section{Summary}
Eight classification methods (GP, KNN, RF, SVC, DT, GNB, ESC, and MKClASS) were compared in this study. The RF algorithm is chosen to search for Am stars among early-type stars in LAMOST DR5 and 15,269 Am stars candidate are obtained. We analyzed the top 50 classification features given by the RF classifier,  the total importance of which reached 57.57\%. We recognized these spectral lines that RF classification depends on.  These lines mostly are iron elements and were used to identify Am stars in the step of manual inspection. In addition, we also compared the difference between Am and Ap stars, and labeled Ap candidates in the final catalog.  Finally, we found 9,372 Am stars and 1,131 Ap candidates,  and provided an Am star catalog. We performed statistical analysis of the temperature distribution, spatial distribution, and infrared photometry for these Am stars. The distribution of effective temperature shows that Am stars are mainly concentrated between F0 and A4, with a peak near A7, which is consistent with previous works. The spatial distribution suggests that the frequency of occurrence of Am stars is inversely related to the vertical distance from the Galactic plane ($|\text{Z}|$). We also conducted an infrared photometric study for Am stars. We noticed that the incidence of infrared excess in Am stars gradually reduces from the near-infrared to  mid-infrared range.
 
\acknowledgments
We would like to thank the referee for his/her valuable comments. This work is supported by the National Basic Research Program of China (973 Program, 2014CB845700), China Scholarship Council, the National Natural Science Foundation of China (Grant NO.11390371) and the Joint Research Fund in Astronomy (U1531119) under cooperative agreement between the National Natural Science Foundation of China and Chinese Academy of Sciences. 

Guoshoujing Telescope (the Large Sky Area Multi-Object Fiber Spectroscopic Telescope, LAMOST) is a National Major Scientific Project built by the Chinese Academy of Sciences. Funding for the project has been provided by the National Development and Reform Commission. LAMOST is operated and managed by the National Astronomical Observatories, Chinese Academy of Sciences.

\appendix

\section{The Catalog of Am Stars}
The catalog has 10,503 entries including 9,372 Am stars and 1,131 Ap candidates. Only the top 10 entries are shown as examples. The complete Am  catalog can be downloaded from http://paperdata.china-vo.org/Qinli/2018/dr5\_Am.csv. The first  column show the name of FITS file of each spectrum. The next two columns are right ascension, and declination of J2000 in degrees. $T_{\rm eff}$ is the effective temperature obtained through matching with the Kurucz grid in the [3900\AA, 5600\AA] wavelength range. Fe\_EW and K\_EW are the equivalent widths of the Fe-group lines and CaII K line in the observed spectrum, respectively. Fe\_EW\_m and K\_EW\_m are the equivalent widths of the Fe-group lines and CaII K line of the corresponding Kurucz template, respectively.  The smaller the value of K\_EW - K\_EW\_m and the larger the value of Fe\_EW - Fe\_EW\_m are, the more obvious the Am phenomenon is. K\_type , H\_type , and m\_type are spectral subtypes of the CaII K line, Balmer lines, and metallic lines respectively. The column Z  shows the vertical distance from the Galactic plane. The Ap\_flag is a flag column, Ap\_flag = 1 indicates that the star is an Ap candidate. The r$_{est}$, r$_{lo}$, and r$_{hi}$ come from the catalog in \citet{2018AJ....156...58B} and are the estimated distance,  lower limit distance, and upper limit distances, respectively. Parallax and parallax\_error are taken from Gaia DR2 catalog. The next six columns list the magnitudes and errors of the J, H, and K bands from the 2MASS catalog. The next four columns are  magnitudes and errors of the W1 and W2 bands from the WISE catalog. The $E_{B-V}$ values are taken from \citet{2011ApJ...737..103S}. The next three columns show the color of (J-H), (H-K), and (W1-W2), correcting the dust extinction. The column FeH\_lamost is metal abundance provide by LAMOST pipeline. The last two columns are EW of spectral line of the observed spectra and templates at 4077\AA\,. The greater the value of 4077\_EW - 4077\_EW\_m is, the greater the probability of Ap star is.  
\begin{longrotatetable}
\begin{deluxetable*}{llllllllllll}
\tablecaption{ The Catalog of Am stars in LAMOST DR5 \label{chartable}}
\tablewidth{800pt}
\tabletypesize{\scriptsize}
\tablehead{
\colhead{Name} & \colhead{R.A.(2000)} & \colhead{Dec.(2000)} & \colhead{Teff} & \colhead{Fe\_EW} &  
\colhead{Fe\_EW\_m} & \colhead{K\_EW} & \colhead{K\_EW\_m} & \colhead{K\_type} & \colhead{H\_type} & \colhead{m\_type}  & \colhead{Z}\\ 
\colhead{Ap\_flag} & \colhead{r$_{est}$} & \colhead{r$_{lo}$} & \colhead{r$_{hi}$} & \colhead{parallax} & \colhead{parallax\_error} & \colhead{Jmag} &
\colhead{e\_Jmag} & \colhead{Hmag} & \colhead{e\_Hmag} & \colhead{Kmag}  & \colhead{e\_Kmag} \\
\colhead{} & \colhead{W1mag} & \colhead{e\_W1mag} & 
\colhead{W2mag} & \colhead{e\_W2mag} & \colhead{E$_{B-V}$} &
\colhead{(J-H)$_o$} & \colhead{(H-K)$_o$} & \colhead{(W1-W2)$_o$} & \colhead{FeH\_lamost}& \colhead{4077\_EW} &\colhead{4077\_EW\_m}
} 
\startdata
spec-56994-HD053243N321131V01\_sp06-044 & 85.285 & 31.919 & 7500 & 11.186 & 4.761 & 2.158 & 2.905 & A5V & A7V & A7V & 907.046 \\
0 & 1338.367 & 1290.222 & 1390.170 & 0.719 & 0.028 & 12.304 & 0.021 & 12.108 & 0.023 & 12.001 & 0.023\\
  & 11.909 & 0.022 & 11.112 & 0.022 & 0.962 & -0.054 & -0.047 & 0.772 &\nodata  & 0.717 & 0.830\\ 
spec-56658-GAC096N23B1\_sp10-050 & 94.486 & 22.630 & 7250 & 11.333 & 5.398 & 2.419 & 3.452 & A6IV & A7V & A7V & -88.073\\
0 & 1514.845 & 1415.680 & 1628.588 & 0.633 & 0.045 & 13.052 & 0.026 & 12.887 & 0.033 & 12.775 & 0.027\\
& 12.682 & 0.032 & 12.356 & 0.036 & 1.720 & -0.282 & -0.163 & 0.281 &\nodata & 0.620 & 0.999\\ 
spec-56718-HD152630N280739B01\_sp07-141 & 233.581 & 26.746 & 7500 & 11.409 & 2.875 & 1.326 & 2.003 & A1V & A6V & F5 &-229.736\\
0 & 527.862 & 519.542 & 536.447 & 1.866 & 0.030 & 13.489 & 0.026 & 12.932 & 0.024 & 12.827 & 0.037\\
  & 12.749 & 0.024 & 12.479 & 0.023 & 0.046 & 0.545 & 0.098 & 0.269 & 0.127 &0.347 & 0.569\\
spec-57035-HD064910N271125V01\_sp03-179 & 101.314 & 27.159 & 7750 & 11.183 & 4.160 & 1.778 & 2.444 & A6V & A3IV & A7V &-1700.226\\
0 & 1746.482 & 1655.268 & 1848.073 & 0.543 & 0.031 & 12.221 & 0.020 & 12.170 & 0.028 & 12.101 &0.027\\
& 11.676 & 0.020 & 11.418 & 0.023 & 0.094 & 0.026 & 0.054 & 0.256 &\nodata & 0.672 & 0.691\\
spec-57393-M31025N38B2\_sp13-112 & 27.934 & 39.496 & 7500 & 14.324 & 4.761 & 1.165 & 2.905 & A6V & A7V & A7V &-228.052\\
0 & 2756.721 & 2471.088 & 3109.958 & 0.316 & 0.043 & 14.627 & 0.039 & 14.532 & 0.066 & 14.500 &0.097\\
& 14.327 & 0.028 & 14.109 & 0.043 & 0.033 & 0.086 & 0.027 & 0.217 & 0.458 & 0.678 & 0.830\\
spec-57044-HD020328N513905V01\_sp01-023 & 31.425 & 49.274 & 8250 & 8.412 & 5.101 & 2.009 & 2.777 & A5V & A6IV & A6IV &1350.067\\
0 & 1993.598 & 1848.858 & 2162.054 & 0.470 & 0.039 & 12.858 & 0.023 & 12.860 & 0.023 & 12.794 &0.024\\
& 12.630 & 0.025 & 12.462 & 0.025 & 0.173 & -0.047 & 0.038 & 0.164 & 0.067 &0.402 & 0.786\\
spec-57651-HD005305N571308V01\_sp09-160 & 13.906 & 57.945 & 8250 & 12.201 & 7.701 & 2.496 & 4.074 & A6IV & A6IV & A7V &1121.101\\
0 & 1146.341 & 1098.812 & 1198.080 & 0.845 & 0.037 & 11.811 & 0.026 & 11.704 & 0.030 & 11.640 &0.021\\
& 11.572 & 0.023 & 11.406 & 0.021 & 0.329 & 0.021 & 0.011 & 0.157&\nodata &0.614 & 1.175\\
spec-57328-HD020325N544136B01\_sp16-175 & 29.186 & 55.993 & 7500 & 15.029 & 4.761 & 1.551 & 2.905 & A5V & A7V & A7V & 2265.005\\
1 & 4253.456 & 3630.378 & 5093.715 & 0.186 & 0.041 & 14.563 & 0.039 & 14.366 & 0.064 & 14.287 & 0.070\\
 & 14.166 & 0.034 & 14.082 & 0.049 & 0.317 & 0.115 & 0.028 & 0.076&\nodata &0.863& 0.830\\
spec-57396-GAC073N15B1\_sp08-007s & 74.779 & 14.537 & 8000 & 10.622 & 3.601 & 0.985 & 1.997 & A1IV & A6IV & A7V & 2448.776\\
0 & 2614.076 & 2334.988 & 2962.641 & 0.340 & 0.046 & 14.316 & 0.027 & 14.123 & 0.038 & 14.025 & 0.053\\
 & 13.876 & 0.030 & 13.725 & 0.043 & 0.348 & 0.102 & 0.042 & 0.142&\nodata &0.524& 0.578\\
 spec-57034-GAC103N38B1\_sp03-062s & 101.347 & 39.148 & 8250 & 9.439 & 3.105 & 0.957 & 1.560 & A3IV & A6IV & A7V & 215.197\\
 0 & 3377.005 & 3036.260 & 3796.369 & 0.254 & 0.034 & 13.951 & 0.024 & 13.910 & 0.032 & 13.930 & 0.050\\
  & 13.820 & 0.029 & 13.706 & 0.047 & 0.105 & 0.014 & -0.037 & 0.111 & 0.260 & 0.337 & 0.482 \\
\enddata
\end{deluxetable*}
\end{longrotatetable}


\bibliography{reference}

\begin{thebibliography}{}
\expandafter\ifx\csname natexlab\endcsname\relax\def\natexlab#1{#1}\fi
\providecommand{\url}[1]{\href{#1}{#1}}
\providecommand{\dodoi}[1]{doi:~\href{http://doi.org/#1}{\nolinkurl{#1}}}
\providecommand{\doeprint}[1]{\href{http://ascl.net/#1}{\nolinkurl{http://ascl.net/#1}}}
\providecommand{\doarXiv}[1]{\href{https://arxiv.org/abs/#1}{\nolinkurl{https://arxiv.org/abs/#1}}}

\bibitem[{{Abt}(1981)}]{1981ApJS...45..437A}
{Abt}, H.~A. 1981, \apjs, 45, 437, \dodoi{10.1086/190719}

\bibitem[{{Abt}(2017)}]{2017PASP..129d4201A}
---. 2017, \pasp, 129, 044201, \dodoi{10.1088/1538-3873/aa5b18}

\bibitem[{{Adelman}(1994)}]{1994MNRAS.271..355A}
{Adelman}, S.~J. 1994, \mnras, 271, 355, \dodoi{10.1093/mnras/271.2.355}

\bibitem[{{Bailer-Jones} {et~al.}(2018){Bailer-Jones}, {Rybizki}, {Fouesneau},
  {Mantelet}, \& {Andrae}}]{2018AJ....156...58B}
{Bailer-Jones}, C.~A.~L., {Rybizki}, J., {Fouesneau}, M., {Mantelet}, G., \&
  {Andrae}, R. 2018, \aj, 156, 58, \dodoi{10.3847/1538-3881/aacb21}

\bibitem[{{Balona} {et~al.}(2015){Balona}, {Catanzaro}, {Abedigamba}, {Ripepi},
  \& {Smalley}}]{2015MNRAS.448.1378B}
{Balona}, L.~A., {Catanzaro}, G., {Abedigamba}, O.~P., {Ripepi}, V., \&
  {Smalley}, B. 2015, \mnras, 448, 1378, \dodoi{10.1093/mnras/stv076}

\bibitem[{{Burkhart} \& {Coupry}(2000)}]{2000AA...354..216B}
{Burkhart}, C., \& {Coupry}, M.~F. 2000, A\&A, 354, 216

\bibitem[{{Castelli} \& {Kurucz}(2003)}]{2003IAUS..210P.A20C}
{Castelli}, F., \& {Kurucz}, R.~L. 2003, in IAU Symposium, Vol. 210, Modelling
  of Stellar Atmospheres, ed. N.~{Piskunov}, W.~W. {Weiss}, \& D.~F. {Gray},
  A20

\bibitem[{{Catanzaro} \& {Ripepi}(2014)}]{2014MNRAS.441.1669C}
{Catanzaro}, G., \& {Ripepi}, V. 2014, \mnras, 441, 1669,
  \dodoi{10.1093/mnras/stu674}

\bibitem[{{Catanzaro} {et~al.}(2015){Catanzaro}, {Ripepi}, {Biazzo},
  {Bus{\'a}}, {Frasca}, {Leone}, {Giarrusso}, {Munari}, \&
  {Scuderi}}]{2015MNRAS.451..184C}
{Catanzaro}, G., {Ripepi}, V., {Biazzo}, K., {et~al.} 2015, \mnras, 451, 184,
  \dodoi{10.1093/mnras/stv952}

\bibitem[{{Chen} {et~al.}(2017){Chen}, {Liu}, \& {Shan}}]{2017AJ....153..218C}
{Chen}, P.~S., {Liu}, J.~Y., \& {Shan}, H.~G. 2017, \aj, 153, 218,
  \dodoi{10.3847/1538-3881/aa679a}

\bibitem[{Conti(1970)}]{1970PASP...82..781C}
Conti, P.~S. 1970, PASP, 82, 781, \dodoi{10.1086/128965}

\bibitem[{{Coupry} {et~al.}(1986){Coupry}, {vant Veer-Menneret}, \&
  {Burkhart}}]{1986AAS...64..477C}
{Coupry}, M.~F., {vant Veer-Menneret}, C., \& {Burkhart}, C. 1986, A\&AS, 64,
  477

\bibitem[{{Cui} {et~al.}(2012){Cui}, {Zhao}, {Chu}, {Li}, {Li}, {Zhang}, {Su},
  {Yao}, {Wang}, {Xing}, {Li}, {Zhu}, {Wang}, {Gu}, {Luo}, {Xu}, \&
  {Zhang}}]{2012RAA....12.1197C}
{Cui}, X.-Q., {Zhao}, Y.-H., {Chu}, Y.-Q., {et~al.} 2012, Research in Astronomy
  and Astrophysics, 12, 1197, \dodoi{10.1088/1674-4527/12/9/003}

\bibitem[{{Fossati} {et~al.}(2008{\natexlab{a}}){Fossati}, {Bagnulo},
  {Landstreet}, {Wade}, {Kochukhov}, {Monier}, {Weiss}, \&
  {Gebran}}]{2008AA...483..891F}
{Fossati}, L., {Bagnulo}, S., {Landstreet}, J., {et~al.} 2008{\natexlab{a}},
  A\&A, 483, 891, \dodoi{10.1051/0004-6361:200809467}

\bibitem[{{Fossati} {et~al.}(2007){Fossati}, {Bagnulo}, {Monier}, {Khan},
  {Kochukhov}, {Landstreet}, {Wade}, \& {Weiss}}]{2007AA...476..911F}
{Fossati}, L., {Bagnulo}, S., {Monier}, R., {et~al.} 2007, A\&A, 476, 911,
  \dodoi{10.1051/0004-6361:20078320}

\bibitem[{{Fossati} {et~al.}(2008{\natexlab{b}}){Fossati}, {Bagnulo}, {Monier},
  {Khan}, {Kochukhov}, {Landstreet}, {Wade}, \& {Weiss}}]{2008CoSka..38..123F}
---. 2008{\natexlab{b}}, Contributions of the Astronomical Observatory Skalnate
  Pleso, 38, 123

\bibitem[{{Fossati} {et~al.}(2008{\natexlab{c}}){Fossati}, {Kolenberg},
  {Reegen}, \& {Weiss}}]{2008AA...485..257F}
{Fossati}, L., {Kolenberg}, K., {Reegen}, P., \& {Weiss}, W.
  2008{\natexlab{c}}, A\&A, 485, 257, \dodoi{10.1051/0004-6361:200809541}

\bibitem[{{Gebran} {et~al.}(2008){Gebran}, {Monier}, \&
  {Richard}}]{2008AA...479..189G}
{Gebran}, M., {Monier}, R., \& {Richard}, O. 2008, A\&A, 479, 189,
  \dodoi{10.1051/0004-6361:20078807}

\bibitem[{{Gray} \& {Corbally}(2009)}]{2009ssc..book.....G}
{Gray}, R.~O., \& {Corbally}, J., C. 2009, {Stellar Spectral Classification}

\bibitem[{{Gray} {et~al.}(2016){Gray}, {Corbally}, {De Cat}, {Fu}, {Ren},
  {Shi}, {Luo}, {Zhang}, {Wu}, {Cao}, {Li}, {Zhang}, {Hou}, \&
  {Wang}}]{2016AJ....151...13G}
{Gray}, R.~O., {Corbally}, C.~J., {De Cat}, P., {et~al.} 2016, \aj, 151, 13,
  \dodoi{10.3847/0004-6256/151/1/13}

\bibitem[{{Hou} {et~al.}(2015){Hou}, {Luo}, {Yang}, {Wei}, {Zhao}, {Zuo},
  {Song}, {Du}, {Bai}, {Zhang}, {Hou}, \& {Liu}}]{2015MNRAS.449.1401H}
{Hou}, W., {Luo}, A., {Yang}, H., {et~al.} 2015, \mnras, 449, 1401,
  \dodoi{10.1093/mnras/stv176}

\bibitem[{{Iliev} {et~al.}(2006){Iliev}, {Budaj}, {Fe{\v n}ov{\v c}{\'{\i}}k},
  {Stateva}, \& {Richards}}]{2006MNRAS.370..819I}
{Iliev}, I.~K., {Budaj}, J., {Fe{\v n}ov{\v c}{\'{\i}}k}, M., {Stateva}, I., \&
  {Richards}, M.~T. 2006, \mnras, 370, 819,
  \dodoi{10.1111/j.1365-2966.2006.10513.x}

\bibitem[{{Lane} \& {Lester}(1987)}]{1987ApJS...65..137L}
{Lane}, M.~C., \& {Lester}, J.~B. 1987, \apjs, 65, 137, \dodoi{10.1086/191220}

\bibitem[{{Lee} {et~al.}(2008){Lee}, {Beers}, {Sivarani}, {Allende Prieto},
  {Koesterke}, {Wilhelm}, {Re Fiorentin}, {Bailer-Jones}, {Norris}, {Rockosi},
  {Yanny}, {Newberg}, {Covey}, {Zhang}, \& {Luo}}]{2008AJ....136.2022L}
{Lee}, Y.~S., {Beers}, T.~C., {Sivarani}, T., {et~al.} 2008, \aj, 136, 2022,
  \dodoi{10.1088/0004-6256/136/5/2022}

\bibitem[{{Liu} {et~al.}(2015){Liu}, {Cui}, {Zhang}, {Wan}, {Deng}, {Hou},
  {Wang}, {Yang}, \& {Zhang}}]{2015RAA....15.1137L}
{Liu}, C., {Cui}, W.-Y., {Zhang}, B., {et~al.} 2015, Research in Astronomy and
  Astrophysics, 15, 1137, \dodoi{10.1088/1674-4527/15/8/004}

\bibitem[{{Luo} {et~al.}(2012){Luo}, {Zhang}, {Zhao}, {Zhao}, {Cui}, {Li},
  {Chu}, {Shi}, {Wang}, {Zhang}, {Bai}, {Chen}, {Wang}, {Guo}, {Chen}, {Du},
  {Kong}, {Lei}, {Li}, {Song}, {Wu}, {Zhang}, {Zhou}, {Zuo}, {Du}, {He}, {Hou},
  {Dong}, {Li}, {Li}, {Li}, {Song}, {Tian}, {Wang}, {Wu}, {Yang}, {Yuan},
  {Cao}, {Chen}, {Chen}, {Chen}, {Chu}, {Feng}, {Gong}, {Gu}, {Hou}, {Huo},
  {Hu}, {Hu}, {Hu}, {Jia}, {Jiang}, {Jiang}, {Jiang}, {Jin}, {Li}, {Li}, {Li},
  {Li}, {Li}, {Liu}, {Liu}, {Liu}, {Lu}, {Lu}, {Luo}, {Mao}, {Men}, {Ni}, {Qi},
  {Qi}, {Shi}, {Su}, {Sun}, {Su}, {Tang}, {Tao}, {Tu}, {Wang}, {Wang}, {Wang},
  {Wang}, {Wang}, {Wang}, {Wang}, {Wang}, {Wang}, {Wang}, {Wang}, {Wang},
  {Wang}, {Wang}, {Wei}, {Xue}, {Xing}, {Xu}, {Xu}, {Xu}, {Yang}, {Yang},
  {Yao}, {Yu}, {Yuan}, {Zhai}, {Zhang}, {Zhang}, {Zhang}, {Zhang}, {Zhang},
  {Zhang}, {Zhao}, {Zhou}, {Zhu}, {Zhu}, \& {Zou}}]{2012RAA....12.1243L}
{Luo}, A.-L., {Zhang}, H.-T., {Zhao}, Y.-H., {et~al.} 2012, Research in
  Astronomy and Astrophysics, 12, 1243, \dodoi{10.1088/1674-4527/12/9/004}

\bibitem[{{Luo} {et~al.}(2015){Luo}, {Zhao}, {Zhao}, {Deng}, {Liu}, {Jing},
  {Wang}, {Zhang}, {Shi}, {Cui}, {Chu}, {Li}, {Bai}, {Wu}, {Cai}, {Cao}, {Cao},
  {Carlin}, {Chen}, {Chen}, {Chen}, {Chen}, {Chen}, {Chen}, {Chen},
  {Christlieb}, {Chu}, {Cui}, {Dong}, {Du}, {Fan}, {Feng}, {Fu}, {Gao}, {Gong},
  {Gu}, {Guo}, {Han}, {He}, {Hou}, {Hou}, {Hou}, {Hu}, {Hu}, {Hu}, {Huo},
  {Jia}, {Jiang}, {Jiang}, {Jiang}, {Jin}, {Kong}, {Kong}, {Lei}, {Li}, {Li},
  {Li}, {Li}, {Li}, {Li}, {Li}, {Li}, {Li}, {Li}, {Li}, {Li}, {Liang}, {Lin},
  {Liu}, {Liu}, {Liu}, {Liu}, {Lu}, {Luo}, {Mao}, {Newberg}, {Ni}, {Qi}, {Qi},
  {Shen}, {Shi}, {Song}, {Song}, {Su}, {Su}, {Tang}, {Tao}, {Tian}, {Wang},
  {Wang}, {Wang}, {Wang}, {Wang}, {Wang}, {Wang}, {Wang}, {Wang}, {Wang},
  {Wang}, {Wang}, {Wang}, {Wang}, {Wang}, {Wang}, {Wang}, {Wang}, {Wang},
  {Wang}, {Wei}, {Wei}, {Wu}, {Wu}, {Wu}, {Wu}, {Xing}, {Xu}, {Xu}, {Xu},
  {Yan}, {Yang}, {Yang}, {Yang}, {Yang}, {Yao}, {Yu}, {Yuan}, {Yuan}, {Yuan},
  {Yuan}, {Zhai}, {Zhang}, {Zhang}, {Zhang}, {Zhang}, {Zhang}, {Zhang},
  {Zhang}, {Zhang}, {Zhao}, {Zhou}, {Zhou}, {Zhu}, {Zhu}, {Zou}, \&
  {Zuo}}]{2015RAA....15.1095L}
{Luo}, A.-L., {Zhao}, Y.-H., {Zhao}, G., {et~al.} 2015, Research in Astronomy
  and Astrophysics, 15, 1095, \dodoi{10.1088/1674-4527/15/8/002}

\bibitem[{{Monier} \& {Richard}(2004)}]{2004IAUS..224..209M}
{Monier}, R., \& {Richard}, O. 2004, in IAU Symposium, Vol. 224, The A-Star
  Puzzle, ed. J.~{Zverko}, J.~{Ziznovsky}, S.~J. {Adelman}, \& W.~W. {Weiss},
  209--214

\bibitem[{{Morgan} {et~al.}(1978){Morgan}, {Abt}, \&
  {Tapscott}}]{1978rmsa.book.....M}
{Morgan}, W.~W., {Abt}, H.~A., \& {Tapscott}, J.~W. 1978, {Revised MK Spectral
  Atlas for stars earlier than the sun}

\bibitem[{{Przybilla} {et~al.}(2017){Przybilla}, {Aschenbrenner}, \&
  {Buder}}]{2017AA...604L...9P}
{Przybilla}, N., {Aschenbrenner}, P., \& {Buder}, S. 2017, A\&A, 604, L9,
  \dodoi{10.1051/0004-6361/201731384}

\bibitem[{{Renson} {et~al.}(1991){Renson}, {Gerbaldi}, \&
  {Catalano}}]{1991AAS...89..429R}
{Renson}, P., {Gerbaldi}, M., \& {Catalano}, F.~A. 1991, A\&AS, 89, 429

\bibitem[{{Renson} \& {Manfroid}(2009)}]{2009AA...498..961R}
{Renson}, P., \& {Manfroid}, J. 2009, A\&A, 498, 961,
  \dodoi{10.1051/0004-6361/200810788}

\bibitem[{{Roman} {et~al.}(1948){Roman}, {Morgan}, \&
  {Eggen}}]{1948ApJ...107..107R}
{Roman}, N.~G., {Morgan}, W.~W., \& {Eggen}, O.~J. 1948, \apj, 107, 107,
  \dodoi{10.1086/144995}

\bibitem[{{Romanyuk}(2007)}]{2007AstBu..62...62R}
{Romanyuk}, I.~I. 2007, Astrophysical Bulletin, 62, 62,
  \dodoi{10.1134/S1990341307010063}

\bibitem[{{Schlafly} \& {Finkbeiner}(2011)}]{2011ApJ...737..103S}
{Schlafly}, E.~F., \& {Finkbeiner}, D.~P. 2011, \apj, 737, 103,
  \dodoi{10.1088/0004-637X/737/2/103}

\bibitem[{{Skrutskie} {et~al.}(2006){Skrutskie}, {Cutri}, {Stiening},
  {Weinberg}, {Schneider}, {Carpenter}, {Beichman}, {Capps}, {Chester},
  {Elias}, {Huchra}, {Liebert}, {Lonsdale}, {Monet}, {Price}, {Seitzer},
  {Jarrett}, {Kirkpatrick}, {Gizis}, {Howard}, {Evans}, {Fowler}, {Fullmer},
  {Hurt}, {Light}, {Kopan}, {Marsh}, {McCallon}, {Tam}, {Van Dyk}, \&
  {Wheelock}}]{2006AJ....131.1163S}
{Skrutskie}, M.~F., {Cutri}, R.~M., {Stiening}, R., {et~al.} 2006, \aj, 131,
  1163, \dodoi{10.1086/498708}

\bibitem[{{Smalley} {et~al.}(2017){Smalley}, {Antoci}, {Holdsworth}, {Kurtz},
  {Murphy}, {De Cat}, {Anderson}, {Catanzaro}, {Cameron}, {Hellier}, {Maxted},
  {Norton}, {Pollacco}, {Ripepi}, {West}, \& {Wheatley}}]{2017MNRAS.465.2662S}
{Smalley}, B., {Antoci}, V., {Holdsworth}, D.~L., {et~al.} 2017, \mnras, 465,
  2662, \dodoi{10.1093/mnras/stw2903}

\bibitem[{{Smith}(1996)}]{1996ApSS.237...77S}
{Smith}, K.~C. 1996, Ap\&SS, 237, 77, \dodoi{10.1007/BF02424427}

\bibitem[{{Smith}(1971)}]{1971AJ.....76..896S}
{Smith}, M.~A. 1971, AJ, 76, 896, \dodoi{10.1086/111198}

\bibitem[{{Smith}(1973)}]{1973ApJS...25..277S}
---. 1973, \apjs, 25, 277, \dodoi{10.1086/190270}

\bibitem[{{Titus} \& {Morgan}(1940)}]{1940ApJ....92..256T}
{Titus}, J., \& {Morgan}, W.~W. 1940, ApJ, 92, 256, \dodoi{10.1086/144215}

\bibitem[{{Wright} {et~al.}(2010){Wright}, {Eisenhardt}, {Mainzer}, {Ressler},
  {Cutri}, {Jarrett}, {Kirkpatrick}, {Padgett}, {McMillan}, {Skrutskie},
  {Stanford}, {Cohen}, {Walker}, {Mather}, {Leisawitz}, {Gautier}, {McLean},
  {Benford}, {Lonsdale}, {Blain}, {Mendez}, {Irace}, {Duval}, {Liu}, {Royer},
  {Heinrichsen}, {Howard}, {Shannon}, {Kendall}, {Walsh}, {Larsen}, {Cardon},
  {Schick}, {Schwalm}, {Abid}, {Fabinsky}, {Naes}, \&
  {Tsai}}]{2010AJ....140.1868W}
{Wright}, E.~L., {Eisenhardt}, P.~R.~M., {Mainzer}, A.~K., {et~al.} 2010, \aj,
  140, 1868, \dodoi{10.1088/0004-6256/140/6/1868}

\bibitem[{{Yuan} {et~al.}(2013){Yuan}, {Liu}, \& {Xiang}}]{2013MNRAS.430.2188Y}
{Yuan}, H.~B., {Liu}, X.~W., \& {Xiang}, M.~S. 2013, \mnras, 430, 2188,
  \dodoi{10.1093/mnras/stt039}

\bibitem[{{Zhao} {et~al.}(2012){Zhao}, {Zhao}, {Chu}, {Jing}, \&
  {Deng}}]{2012RAA....12..723Z}
{Zhao}, G., {Zhao}, Y.-H., {Chu}, Y.-Q., {Jing}, Y.-P., \& {Deng}, L.-C. 2012,
  Research in Astronomy and Astrophysics, 12, 723,
  \dodoi{10.1088/1674-4527/12/7/002}

\end{thebibliography}

\end{document}